\documentclass[8pt]{iopart}
\usepackage{iopams}
\usepackage{graphicx}
\usepackage{color}
\usepackage{subfig}
\usepackage[titletoc,title]{appendix}
\usepackage{citesort}
\usepackage{footmisc}
\newcommand{\cG}{{\mathbf{\cal G}}}
\newcommand{\bS}{{\mathbf{S}}}
\newcommand{\bM}{{\mathbf{M}}}

\newcommand{\bg}{{\mathbf{g}}}

\newcommand{\besse}{{\mathbf{s}}}

\newcommand{\bJ}{{\mathbf{J}}}
\newcommand{\bh}{{\mathbf{h}}}

\newcommand{\blam}{{\boldsymbol \lambda}}
\newcommand{\bgam}{{\boldsymbol \gamma}}
\newcommand{\be}{{\boldsymbol \eta}}

\begin{document}

\title[]{Variational perturbation and extended Plefka approaches to dynamics on random networks: the case of the kinetic Ising model}

\author{L Bachschmid-Romano\footnote{\label{note1} These authors contributed equally to this work.}$^{1}$, C Battistin \footref{note1} $^{2}$, M Opper$^1$, Y Roudi$^{2,3}$}

\address{$^1$ Department of Artificial Intelligence, Technische Universität Berlin, Marchstra\ss e 23, Berlin 10587, Germany}
\address{$^2$ Kavli Institute and Centre for Neural Computation Trondheim, Norway}
\address{$^3$ Institute for Advanced Study, Princeton,  USA}

\eads{claudia.battistin@ntnu.no, ludovica.bachschmidromano@tu-berlin.de}

\vspace{10pt}

%
%
%
%
%
\begin{abstract}
We describe and analyze some novel approaches for studying the dynamics of Ising spin glass models. We first briefly consider the variational approach based on minimizing the Kullback-Leibler divergence between independent trajectories and the real ones and note that this approach only coincides with the mean field equations from the saddle point approximation to the generating functional when the dynamics is defined through a logistic link function, which is the case for the kinetic Ising model with parallel update. We then spend the rest of the paper developing two ways of going beyond the saddle point approximation to the generating functional. In the first one, we develop a variational perturbative approximation to the generating functional by expanding the action around a quadratic function of the local fields and conjugate local fields whose parameters are optimized. We derive analytical expressions for the optimal parameters and show that when the optimization is suitably restricted, we recover the mean field equations that are exact for the fully asymmetric random couplings \cite{mezard2011exact}. However, without this restriction the results are different.  We also describe an extended Plefka expansion in which in addition to the magnetization, we also fix the correlation and response functions. Finally, we numerically study the performance of these approximations for Sherrington-Kirkpatrick type couplings for various  coupling strengths and the degrees of coupling symmetry, for both temporally constant but random, as well as time varying external fields. We show that the dynamical equations derived from the extended Plefka expansion outperform the others in all regimes, although it is computationally more demanding. The unconstrained variational approach does not perform well in the small coupling regime, while it approaches dynamical TAP equations of \cite{roudi2011dynamical} for strong couplings.
\end{abstract}

\section{Introduction}
The kinetic Ising spin glass model is a prototypical model for studying the dynamics of disordered systems. Previous work on this topic focused both on studying the average -- over couplings --  behavior of various order parameters, such as magnetizations, correlations and response functions, and in more recent years, developing approximate methods for relating the dynamics of a given realization of the model to its parameters. The latter line of work has received a lot of attention in recent years, in part, because of the applications it has on developing approximate inference methods for point processes which in turn are receiving particular attention due to the on going improvements in data acquisition techniques in various disciplines in life sciences. 

Most of the early work on the topic dealt with systems with symmetric interactions, until Crisanti and Sompolinsky \cite{crisanti1987dynamics} studied the disorder averaged dynamics of Ising models with various degrees of symmetry and Kappen and Spanjers \cite{kappen2000mean} derived naive mean field and TAP equations for the stationary state of the Ising model for arbitrary couplings, in both cases considering Glauber dynamics. Roudi and Hertz \cite{roudi2011dynamical} derived dynamical TAP equations 
(hereafter denoted by RH-TAP) for both discrete time parallel and continuous time Glauber dynamics using Plefka's method \cite{plefka1982convergence}, originally used for studying equilibrium spin glass models, extended to dynamics. This was followed by \cite{aurell2012dynamic} who reported another derivation of these equations using information geometry following the approach of \cite{kappen2000mean}. 
Mezard and Sakellariou \cite{mezard2011exact} developed a mean field method (hereafter denoted by MS-MF) which is exact for large networks with independent random couplings and an elegant generalized mean field methods was followed in \cite{mahmoudi2014generalized}. 

In the current paper we follow up on these efforts and report some new results on the dynamics of kinetic Ising model with parallel dynamics. We first look at the relationship between the saddle point approximation to the path integral representation of the dynamics and the simplest variational approach based on minimizing the Kullback-Leibler (KL) divergence between the true distribution of the spin trajectories and a factorized distribution. Although for the standard kinetic Ising model the two methods yield the same equations of motion, we see that this is not in general the case when the probability of spin configurations at a given time given those of the previous time is not a logistic function of the fields. After this, we consider two approaches for going beyond the saddle point solution of the path integral representation of the dynamics of the standard kinetic Ising model with parallel dynamics (defined in more detail in the following sections). 

In one of these approaches, which we refer to as Gaussian Average variational method, we perform a Taylor expansion of the action in the path integral representation of the generating functional around a quadratic function of the fields and conjugate fields. As described in described in detail in section \eref{GaussianAverage}, we then choose the parameters of this function such that the resulting functional minimally depends on these parameters. We derive analytical expressions for these optimal solutions and show that for a fully asymmetric network under a further assumption about the interaction between the fields and the conjugate fields, we can recover the equations of motion for the magnetization identical to MS-MF equations \cite{mezard2011exact}. Without this assumption we observe that the resulting equations are different from MS-MF. In the second approach, we go beyond the saddle point by performing an extended Plefka expansion. The standard Plefka expansion for the equilibrium model involves performing a small coupling approximation of the free energy at fixed magnetization and is the approach that was originally taken in \cite{roudi2011dynamical}.  As we show here, however, similar to the soft spin models \cite{biroli1999dynamical,bravi2015extended}, a better description of the dynamics can be achieved by not only fixing the magnetizations but also pairwise correlation and response functions while expanding around the uncoupled model.

\section{The dynamical model}

We consider the synchronous dynamics of $N$ interacting binary spins in the time window $[0,T]$ defined by 
\begin{equation}
{\rm P}( \besse^{0:T})=\prod_{t=0}^{T-1} {\rm P}(\besse(t+1)\vert \besse(t)){\rm P}\left( \besse(0)\right),
\label{eq:P_distrib}
\end{equation}
in which
\begin{equation}
{\rm P}(\besse(t+1)\vert \besse(t))=\prod_{i=1}^N f_{s_i(t+1)} \left( H_i(t) \right)
\label{eq:PModel0}
\end{equation}
where 
\begin{equation}H_i(t)=h_i(t)+\sum_{j=1}^N J_{ij}s_j(t)
\end{equation}
 is the total field acting on spin $i$ at time $t$ composed of the external field $h_i$ and the fields felt from other spins in the system. The function $f_{s_i(t+1)} \left( H_i(t) \right)$ is a generic transfer function or conditional probability of the state of the spin $i$ at time $(t+1)$ given the field at time $t$. Our goal will be to calculate the mean magnetizations of the spins. 

The generating functional of the distribution ${\rm P}( \besse^{0:T})$, expressed as a path integral integral form is
\begin{equation}
\fl Z\left[ \bpsi,\bh, \bJ\right]=\left\langle e^{i\bpsi^{\top}\besse}\right\rangle_P =\frac{1}{2^N(2\pi)^{NT}}\int  D\mathcal{G} e^{-\mathit{L[\bh,\bpsi,\cG]}},
\label{eq:GenFunc1}  
\end{equation}
where $\left\langle \dots \right\rangle_P$ denotes averaging with respect to the history of trajectories defined by (\ref{eq:P_distrib}) and (\ref{eq:PModel0}), and
\begin{eqnarray}
\mathit{L[\bh,\bpsi,\cG]}\equiv& - i\sum_{i=1}^N \sum_{t=0}^{T-1} \hat{g}_i(t)\left( g_i(t)-h_i(t)\right)\label{eq:GenFunc2}\\
&+\sum_{i=1}^N \sum_{t=0}^{T}\ln {\rm Tr}_{s_i(t)}f_{s_i(t)} \left( g_i(t-1)\right)e^{is_i(t)\left[ \psi_i(t)-\sum_jJ_{ji}\hat{g}_j(t)\right] } \nonumber
\end{eqnarray}
having set $\hat{g}_i(T)=0$ and $f_{s_i(0)}(g_i(-1))=1$.
Notice that we assumed the initial state $\besse(0)$ to be uniformly distributed, 
manifested in the factor $1/2^N$ in (\ref{eq:GenFunc1}),
and that we 
refer to the two auxiliary variables with the compact notation
$\mathcal{G} \equiv \{ g_i(t), \hat{g}_i(t) \}_{i=1 \dots N,\, t=0 \dots T}$
and $ D\mathcal{G} = \prod_{i,t} dg_i(t) d \hat{g}_i(t)$. 

The magnetization of spin $i$ at time $t$ can then be obtained as the first derivative of the log-generating functional:
\begin{equation}
m_i(t)= -i\lim_{\bpsi\rightarrow \mathbf{0}} \frac{\partial \ln (Z\left[ \bpsi,\bh, \bJ\right])}{\partial \psi_i(t)}\label{eq:magn0}.
\end{equation}

Let us make a brief note on how the the integral representation of the generating functional in (\ref{eq:GenFunc1})-(\ref{eq:GenFunc2}) has been derived. This is done by first replacing $H_i(t)$ in \eref{eq:PModel0} by $g_i(t)$ and integrating over all $g_i(t)$ while enforcing that at each time step and for each spin $g_i(t)=H_i(t)$ by inserting $\delta$-functions, $\delta[g_i(t)-H_i(t)]$, in the integral. One then writes this delta function in its the integral representation

\begin{equation}
\delta[g_i(t)-H_i(t)]= \int \frac{d\hat{g}_i(t)}{2\pi} \exp\left \{i\hat{g}_i(t)[g_i(t)-H_i(t)]\right\}
\end{equation}
which is how the $\hat{g}_i(t)$ appear in the equations. This rewriting of the generating functional constitutes the first steps in the Martin-Siggia-Rose-De Domenicis-Peliti formalism \cite{martin1973statistical,de1978field} once it is adapted for hard spins. For more details about this approach and a pedagogical review on its application to soft and hard spin dynamics see \cite{coolen2000} and \cite{hertz2016path}.

A logistic transfer function $f$ in (\ref{eq:PModel0}), such that $f_{s_i(t+1)}(H_i(t))=\frac{1}{2}(1+s_i(t+1)\tanh H_i(t))$, yielding the following probability distribution over spin paths
\begin{equation}
{\rm P}( \besse^{0:T})=\frac{1}{2^N} \prod_{i=1}^N\prod_{t=0}^{T-1}\frac{e^{s_i(t+1)H_i(t)}}{2\cosh(H_i(t))}
\label{eq:PIsing0} \qquad .
\end{equation}
corresponds to the {\em standard} kinetic Ising model with parallel update studied in previous work \cite{roudi2011dynamical,aurell2012dynamic,mezard2011exact,mahmoudi2014generalized}.

This path integral representation in \eref{eq:GenFunc1} allows us to explicitly perform the trace over the spins in the generating functional of (\ref{eq:GenFunc1})-(\ref{eq:GenFunc2}) yielding
\begin{eqnarray}
\fl \mathit{L[\bh,\bpsi,\cG]}\equiv &- \sum_{i=1}^{N}\sum_{t=0}^{T-1} lc \left[  g_i(t-1)+i\psi_i(t)-i\sum_{l}J_{li}\hat{g}_l(t)\right]  \nonumber\\
& -i\sum_{i=1}^{N}\sum_{t=0}^{T-1}\hat{g}_i(t)\left( g_i(t)-h_i(t)\right) \nonumber\\ 
&-\sum_{i=1}^{N} lc \left[ g_i(T-1)+i\psi_i(T)\right]+\sum_{i=1}^{N}\sum_{t=0}^{T-1} lc \left[ g_i(t)\right] 
\label{eq:GenFuncIsing}
\end{eqnarray}
where we have set $g_i(-1)=0$ $\forall i$ and 

\begin{equation}
lc[x]\equiv \log \cosh(x)\nonumber.
\end{equation}

\section{Mean Field}
 \label{sec:MeanField}
As a prologue to our more important results in the following sections, in this section we review the derivation of mean field equations for the dynamical model in (\ref{eq:PModel0}) using two approaches. These are the saddle point approximation to the path integral representation of the generating functional in \eref{eq:GenFunc1}, and the minimization of the KL distance between the true distribution ${\rm P}$ in \eref{eq:P_distrib} and a factorized one. Despite being formally different methods, in the literature they are both often referred to as {\em mean field} and it is indeed well know that for the specific case of the equilibrium Ising model, they lead to the very same set of equations, known as {\em na\"{i}ve mean field equations} \cite{opper2001advanced}. Throughout this section the transfer function $f$ in \eref{eq:PModel0} is considered a generic function of the field $H_i(t)$. Only towards the end of this section we are going to consider $f$ as a logistics function of the kinetic Ising model.

\subsection{Saddle point mean field}
In the equilibrium case, one way to derive the na\"{i}ve mean field equations is as the equations describing the saddle point approximation to a path integral representation of the free energy, while the TAP equations are those derived by calculating the Gaussian integral around the saddle point \cite{kholodenko1990onsager}. (Another way is by means of Plefka expansion, which at this point we do not discuss but will get back to later on ). Let us consider this saddle point approach for the kinetic model in (\ref{eq:PModel0}) and the corresponding generating functional (\ref{eq:GenFunc1}). Defining a complex measure $q$ as
\begin{equation}
\fl q\left( s_i(t)\vert g_i(t-1),\hat{\bg}(t),\psi_i(t)\right)=\frac{f_{s_i(t)} \left( g_i(t-1)\right)e^{is_i(t)\left[ \psi_i(t)-\sum_jJ_{ji}\hat{g}_j(t)\right] }}{{\rm F}_{it}\left( g_i(t-1),\hat{\bg}(t),\psi_i(t)\right) }
\label{eq:PMF1a} 
\end{equation}
where ${\rm F}_{it}\left( g_i(t-1),\hat{\bg}(t),\psi_i(t)\right)$ is the normalization constant, the saddle point equations for the generating functional of (\ref{eq:GenFunc1}), namely the stationary points of the function $\mathit{L[\bh,\psi,\cG]}$, in (\ref{eq:GenFunc2}), $\hat{g}^{\rm SP}_i(t)$ and $g_i^{\rm SP}(t)$, read
\numparts
\begin{eqnarray}
\hat{g}^{\rm SP}_i(t)&=&i\left\langle \frac{f^\prime_{s_i(t+1)} \left( g_i^{\rm SP}(t)\right)}{f_{s_i(t+1)} \left( g_i^{\rm SP}(t)\right)}\right\rangle_{q\left( s_i(t+1)\vert g_i^{\rm SP}(t),\hat{\bg}^{\rm SP}(t+1),\psi_i(t+1)\right)}\label{eq:PMF2a} \\ 
g_i^{\rm SP}(t)&=&h_i(t)+\sum_j J_{ij}\left\langle s_j(t)\right\rangle_{q\left( s_j(t)\vert g_j^{\rm SP}(t-1),\hat{\bg}^{\rm SP}(t),\psi_j(t)\right)} \label{eq:PMF2b} 
\end{eqnarray}
\endnumparts
where we have defined $f^\prime_x(y)\equiv \frac{\partial f_x(y)}{\partial y}$.
Notice that in the limit $\bpsi\to \mathbf{0}$, $\hat{\bg}^{\rm SP}=\mathbf{0}$ is a self-consistent solution of the previous saddle point equation (\ref{eq:PMF2a}), while  (\ref{eq:PMF2b}) turns into
\begin{equation}
g_i^{\rm SP}(t)=h_i(t)+\sum_j J_{ij}\left\langle s_j(t)\right\rangle_{f_{s_j(t)}\left(  g_j^{\rm SP}(t-1)\right) } \label{eq:PMF3}
\end{equation} 
The approximate log generating functional $\ln Z\left[ \bpsi,\bh, \bJ\right]\simeq  -\mathit{L[\bh,\bpsi,\cG^{\rm SP}]}+const.$ allows us to estimate the magnetizations using (\ref{eq:magn0}) and (\ref{eq:PMF3}) as
\begin{equation}
m_i(t)=\left\langle s_i(t)\right\rangle_{f_{s_i(t)}\left(  h_i(t-1)+\sum_j J_{ij}m_j(t-1)\right) }.
\label{eq:PMF5}
\end{equation}
These are the saddle point mean field equations for a general function $f$. Note that the marginal here yields the same expression as the conditional probability in \eref{eq:PModel0}, namely $f_{s_i(t+1)}\left(  H_i(t)\right) $ except that in \eref{eq:PMF5}, the fluctuating field $H_i(t)$ has been replaced by an effective (mean) field $H^{\rm eff}_i(t)= h_i(t)+\sum_j J_{ij}m_j(t)$, in analogy with the physical intuition behind the original formulation of the mean field theory by Weiss \cite{weiss1907hypothese}.

\subsection{Mean field from KL distance}
A second way of deriving mean field equations, usually employed in the machine learning community, is based on a variational approximation. Within this framework, one approximates the model distribution ${\rm P}\left( \besse^{0:T}\right) $ with a Markovian process ${\rm Q}\left( \besse^{0:T}\right) $ that factorizes over the spin trajectories \cite{bishop2006pattern}. In other words, assuming
\begin{equation}
{\rm Q}(\besse^{0:T})=\prod_{t=0}^{T-1} {\rm Q}(\besse(t+1)\vert \besse(t)){\rm Q}(\besse(0)),
\label{eq:VMF0}
\end{equation}
where
\begin{equation}
{\rm Q}(\besse(t+1)\vert\besse(t))=\prod_{j=1}^N{\rm Q}(s_j(t+1)\vert s_j(t)),
\label{eq:VMF1b}
\end{equation}
one minimizes the Kullback-Leibler divergence, ${\rm D_{KL}}[{\rm Q}(\besse^{0:T})\|{\rm P}(\besse^{0:T})]$, between the approximate distribution ${\rm Q}(\besse^{0:T})$ and the model  ${\rm P}(\besse^{0:T})$. In the case of the model defined in \eref{eq:PModel0} and an approximate distribution satisfying \eref{eq:VMF0} and \eref{eq:VMF1b}, the KL-divergence can be rewritten as

\numparts
\begin{eqnarray}
\fl & {\rm D_{KL}}[{\rm Q}(\besse^{0:T})\|{\rm P}(\besse^{0:T})] \equiv {\rm Tr}_{\besse^{0:T}}{\rm Q}(\besse^{0:T})\ln \frac{{\rm Q}(\besse^{0:T})}{{\rm P}(\besse^{0:T})} \\
\fl &=\sum_t{\rm Tr}_{\besse(t) }{\rm Q}(\besse(t)){\rm Tr}_{\besse(t+1) }{\rm Q}(\besse(t+1)\vert\besse(t))\ln \frac{{\rm Q}(\besse(t+1)\vert\besse(t))}{{\rm P}(\besse(t+1)\vert\besse(t))}\nonumber \\
\fl &=\sum_{t}{\rm Tr}_{s_j(t) }{\rm Q}(s_j(t)){\rm Tr}_{s_j(t+1) }{\rm Q}(s_j(t+1)\vert s_j(t))\ln\frac{{\rm Q}(s_j(t+1)\vert s_j(t))}{u_{jt}(s_j(t+1)\vert s_j(t))}+\nonumber\\
\fl &\sum_{t}\sum_{i\neq j}{\rm Tr}_{s_i(t) }{\rm Q}(s_i(t)){\rm Tr}_{s_i(t+1) }{\rm Q}(s_i(t+1)\vert s_i(t))\ln{\rm Q}(s_i(t+1)\vert s_i(t)) 
\label{eq:VMF2}
\end{eqnarray}
\endnumparts
where the first line is just the definition of the KL-divergence, in the second line we have exploited the Markovian property of ${\rm P}$ and ${\rm Q}$ and assumed ${\rm P}(\besse(0))={\rm Q}(\besse(0))$, while in the last line we have use the factorizability of ${\rm Q}$ over spin trajectories. Notice that the last equality is valid for any choice of $j$ and that we have defined $u_{jt}$ as
\begin{equation}
\fl u_{jt}(s_j(t+1)\vert s_j(t))\equiv \exp\left\lbrace {\rm Tr}_{\left\lbrace \besse_{\setminus j}(t+1),\besse_{\setminus j}(t)\right\rbrace }{\rm Q}(\besse_{\setminus j}(t+1),\besse_{\setminus j}(t)) \ln {\rm P}(\besse(t+1)\vert\besse(t))\right\rbrace
\label{eq:VMF3}  
\end{equation}
and $\besse_{\setminus j}(t)$ denotes all components of $\besse(t)$ apart from $j$. Observe that thanks to the Markovian property of the two distributions ${\rm P}$ and ${\rm Q}$  we were able to reduce the average over a $NT$ dimensional space to a sum of $T$ averages over $2N$ dimensional spaces.

In order to determine the variational mean field equations, one has to minimize the KL-divergence in the space of marginals ${\rm Q}(s_j(t))$ and transition probabilities ${\rm Q}(s_j(t+1)\vert s_j(t))$. Given that these are not independent, we enforce the constraints:

\begin{equation}
{\rm Q}(s_j(t+1))={\rm Tr}_{s_j(t)}{\rm Q}(s_j(t+1)\vert s_j(t)){\rm Q}(s_j(t))
\label{eq:MarginalQ}
\end{equation}
using Lagrange multipliers $\lambda(s_j(t))$, ultimately optimizing the following cost function:

\begin{eqnarray}
\fl {\cal L}\equiv {\rm D_{KL}}[{\rm Q}(\besse^{0:T})\|{\rm P}(\besse^{0:T})]\nonumber\\
\fl \ \ \ \ -\sum_{j,t}{\rm Tr}_{s_j(t)}\lambda(s_j(t))\left\lbrace {\rm Q}(s_j(t))- {\rm Tr}_{s_j(t-1)}{\rm Q}(s_j(t)\vert s_j(t-1)){\rm Q}(s_j(t-1))\right\rbrace  
\label{eq:VMF4} \qquad . 
\end{eqnarray}
The stationary points of ${\cal L}$ in (\ref{eq:VMF4}) are the zeros of the functional derivatives

\numparts
\begin{eqnarray}
\fl \frac{\delta{\cal L}}{\delta {\rm Q}(s_j(t))}&=&{\rm Tr}_{s_j(t+1)}{\rm Q}(s_j(t+1)\vert s_j(t))\ln \frac{{\rm Q}(s_j(t+1)\vert s_j(t))}{u_{jt}(s_j(t+1)\vert s_j(t))}-\lambda(s_j(t))+\nonumber\\
&&{\rm Tr}_{s_j(t+1)}\lambda(s_j(t+1)){\rm Q}(s_j(t+1)\vert s_j(t))\\
\fl \frac{\delta{\cal L}}{\delta {\rm Q}(s_j(t+1)\vert s_j(t))}&=&{\rm Q}(s_j(t))\left\lbrace \ln \frac{{\rm Q}(s_j(t+1)\vert s_j(t))}{u_{jt}(s_j(t+1)\vert s_j(t))}+1+\lambda(s_j(t+1))\right\rbrace 
\end{eqnarray}
\endnumparts
that can be reduced to the relation:

\begin{equation}
{\rm Q}(s_j(t+1)\vert s_j(t))=\frac{u_{jt}(s_j(t+1)\vert s_j(t))}{{\rm Tr}_{s_j(t+1)}u_{jt}(s_j(t+1)\vert s_j(t))}.
\label{eq:VMF5} 
\end{equation}

It is worth emphasizing that this solution is valid for any Markov chain ${\rm P}$ and any approximate Markov distribution ${\rm Q}$ that factorizes over the spin trajectories. From now on we will require the spins at time $t$ to be conditionally independent under the model distribution, as in (\ref{eq:PModel0}). This assumption and a little algebra allow us to simplify (\ref{eq:VMF5}) as follows:
\begin{equation}
\fl {\rm Q}[s_j(t+1)\vert s_j(t)]=\frac{\exp\left\lbrace {\rm Tr}_{\besse_{\setminus j}(t)} {\rm Q}[\besse_{\setminus j}(t)]\ln f_{s_j(t+1)} \left( h_j(t)+\sum_l J_{jl}s_l(t)\right)\right\rbrace }{
{\rm Tr}_{s_j(t+1)}\exp \left\lbrace {\rm Tr}_{\besse_{\setminus j}(t)} {\rm Q}[\besse_{\setminus j}(t)]\ln f_{s_j(t+1)} \left( h_j(t)+\sum_l J_{jl}s_l(t)\right)\right\rbrace} 
\label{eq:VMF6}
\end{equation}
where we imposed the normalizability to ${\rm Q}$.

If there are no self-couplings in the model distribution ${\rm P}$, the right-hand side of (\ref{eq:VMF6}) will not depend on $s_j(t)$ and consequently the solution for the joint distribution ${\rm Q}(\besse^{0:T})$ will factorize in time. The spin independent 1-st order Markov chain ${\rm Q}$ that best approximates the model ${\rm P}$ defined in (\ref{eq:PModel0}) with $J_{jj}=0$, is actually a 0-th order Markov chain.  Additionally the absence of self-interactions in ${\rm P}$ makes (\ref{eq:VMF6}) an explicit relation between the marginal of spin $j$ at time $t+1$ and the marginals of all spins but $j$ at the previous time step $t$. Since we are dealing with a system of binary units, marginals are fully determined by their first moments, thus the marginal of spin $j$ at time $t+1$, in (\ref{eq:VMF6}), becomes a function of the magnetizations at time $t$. Taking one step further one can easily verify that the first moments of \eref{eq:VMF6} equal the na\"{i}ve mean field 
magnetizations of (\ref{eq:PMF5}) if the transition probability $f_{s_i(t+1)} \left( H_i(t)\right)$ belongs to the exponential family with the field $H_i(t)$ as natural parameter

\begin{equation}
f_{s_i(t+1)}(H_i(t))=\frac{\exp\left[ a(s_i(t+1))H_i(t)\right] }{{\rm Tr}_{s_i(t)}\exp\left[ a(s_i(t+1))H_i(t)\right]},
\label{eq:ExpFam}
\end{equation} 
where $a(\cdot)$ is a generic function of the state $s_i(t+1)$. For the kinetic Ising model $a(\cdot)$ is the identity function and the equations for the magnetizations read:

\begin{equation}
m_i(t)= \tanh\left[ h_i(t-1) + \sum_j J_{ij}m_j(t-1)\right],
\label{eq:NMF} 
\end{equation}
equivalent to (\ref{eq:PMF5}) and know as the dynamical Na\"{i}ve Mean Field equations \cite{roudi2011dynamical}.

\section{Gaussian Average method}\label{GaussianAverage}

What we have shown so far is that the saddle point approximation to the generating functional for the kinetic Ising model and the one based on the KL divergence match each other, although this is not the case for non-logistics transfer functions. In this section, we study an improvement over the saddle point approximation. Our approach is to find the optimal Gaussian distribution for approximating the generating functional perturbatively, and then using the resulting approximation to calculate the magnetizations. This can be thought as an extension to complex measures of a standard variational method: it was taken by M\"{u}schlegel and Zittartz \cite{muhschlegel63gaussian} for the equilibrium Ising model, while a general framework is set in \cite{sissakian1992variational}. We describe this approach in detail in this section.

\subsection{Optimization}
\label{sec:Optimization}

We consider the first order Taylor expansion of the log-generating functional defined in (\ref{eq:GenFunc1})-(\ref{eq:GenFunc2}) around a gaussian integral:
\begin{equation}
\fl -\ln (Z\left[ \bpsi,\bh, \bJ\right])\simeq -\ln \int \tilde{D} \cG  +\frac{\int \tilde{D} \cG \left( L-L_s\right)}{\int \tilde{D} \cG } +NT\ln2\pi +N\ln 2,\label{eq:GaussGenFunc1} 
\end{equation}
where we have defined the complex gaussian measure
\numparts
\begin{eqnarray}
\tilde{D} \cG &=& D \cG e^{-L_s} \label{eq:GaussMeasure}\\
L_s&=&\frac{1}{2}\left(\cG-\bar{\be}\right)^\top \bS\left(\cG-\bar{\be}\right) \label{eq:GaussForm}
\end{eqnarray}
\endnumparts
parametrized by the interaction matrix $\bS$ and the mean $\bar{\be}$. Here we split the vectors $\bar{\be}$ into $\{\be(t),\hat{\be}(t)\}_{t=0}^{T-1}$ and $\be(t)$ into $\{\eta_i(t)\}_{i=1}^N$  similar to $\cG$. From now on we will use the form of the action $L$ in (\ref{eq:GenFuncIsing}) since we are going to focus on the standard parallel update kinetic Ising model.

The choice of a quadratic form for $L_s$ allows us to easily calculate many of the terms in (\ref{eq:GaussGenFunc1}), simplifying the expression for the log-generating functional as
\begin{eqnarray}
\fl -\ln Z= \frac{1}{2}\ln \det \bS -NT -i\sum_{i,t}\bS^{-1}_{2Nt+i;2Nt+N+i}-i\sum_{i,t}\eta_i(t)\hat{\eta}_i(t)+i\sum_{i,t} \hat{\eta}_i(t)h_i(t) \nonumber\\
\fl-\frac{\sqrt{\det \bS}}{\left( 2\pi\right)^{NT}}\sum_{i}\sum_{t=1}^{T-1}\int \tilde{D}\cG'  lc \left[ g'_i(t-1) +\eta_i(t-1)+i\psi_i(t) -i\sum_l J_{li}  \hat{g}'_l(t)-i\sum_l J_{li}\hat{\eta}_l(t)\right]\nonumber \\
\fl-\frac{\sqrt{\det \bS}}{\left( 2\pi\right)^{NT}}\sum_{i}\int \tilde{D}\cG'  lc \left[ i\psi_i(0) -i\sum_l J_{li}  \hat{g}'_l(0)-i\sum_l J_{li}\hat{\eta}_l(0)\right]\nonumber \\
\fl-\frac{\sqrt{\det \bS}}{\left( 2\pi\right)^{NT}}\sum_{i}\int \tilde{D}\cG' lc \left[g'_i(T-1) +\eta_i(T-1)+i\psi_i(T)\right] \nonumber\\
\fl+\frac{\sqrt{\det \bS}}{\left( 2\pi\right)^{NT}}\sum_{i,t}\int \tilde{D}\cG' lc \left[g'_i(t) +\eta_i(t)\right]+N\ln 2 \label{eq:lnZTOT}
\end{eqnarray}
where we have replaced (\ref{eq:GenFuncIsing}) in (\ref{eq:GaussGenFunc1}) and we have performed the change of variables $\cG'=\cG-\bar{\be}$. Notice that when not stated otherwise the sum over $t$ runs from $t=0$ to $t=T-1$. From now on we will just drop off the superscript $'$ from variables $\cG$.

If all measures were real probability measures, the first order approximation
on the right hand side of \eref{eq:GaussGenFunc1} would be an upper bound to the 
free energy $-\ln Z$. In this case a minimization of the bound with respect 
to the variational parameters would be the obvious choice for optimizing
the approximation. Since integrations in our case are over complex measures
this argument cannot be applied. Instead, we base our optimisation
on the idea of the Variational Perturbation method \cite{kleinert2009path}:
if the Taylor series expansion of the log generating functional \eref{eq:GenFunc1}-\eref{eq:GenFuncIsing}  would be continued to infinite order it would represent the functional and the resulting series would be 
entirely independent of the parameters of the gaussian measure (\ref{eq:GaussMeasure}).
On the other hand, the truncated series (\ref{eq:GaussGenFunc1}) inherits a dependence on the variational parameters $\be$, $\hat{\be}$, $\bS$. Hence, one would expect 
that the truncation represents the most sensible approximation
if it depends the least on these parameters. One should therefore choose their optimal values such that the approximation to 
$\ln Z$ is the most insensitive to variations of these parameters. This simply corresponds to computing the stationary values of the log generating functional in the $\be$, $\hat{\be}$, $\bS$ space.  
This requirement of minimum sensitivity to the variational parameters was introduced in \cite{stevenson1981optimized} as an approximation protocol. 

Using the logic in the previous paragraph and setting the first derivative of the expression for $-\ln Z$ in \eref{eq:lnZTOT} with respect to $\hat{\eta}_j(t)$ to zero, one gets the equation for stationary $\eta_j(t)$, the first moment of the gaussian form for $g_j(t)$:

\begin{equation}
\eta_i(t)= h_i(t)+\sum_j J_{ij}\mu_i(t) \label{eq:SaddleAlpha}\\
\end{equation}
where we have defined for $t=1,...,T-1$:

\begin{eqnarray}
\fl \mu_i(t)=\frac{\sqrt{\det \bS}}{\left( 2\pi\right)^{NT}}\int \tilde{D}\cG \tanh \left[ g_i(t-1)+\eta_i(t-1)+i\psi_i(t)-i\sum_l J_{li}(\hat{g}_l(t)+\hat{\eta}_l(t)) \right],\nonumber\\
\ \ \label{eq:DefMus}
\end{eqnarray}
while for $t=0$ and $t=T$ we have respectively:

\numparts
\begin{eqnarray}
\fl \mu_i(0)&=&\frac{\sqrt{\det \bS}}{\left( 2\pi\right)^{NT}}\int \tilde{D}\cG \tanh \left[ i\psi_i(0)-i\sum_l J_{li}\left( \hat{g}_l(0)+\hat{\eta}_l(0)\right) \right]\label{eq:DefMus0}\\
\fl \mu_i(T)&=&\frac{\sqrt{\det \bS}}{\left( 2\pi\right)^{NT}}\int \tilde{D}\cG \tanh \left[ g_i(T-1)+\eta_i(T-1)+i\psi_i(T)\right]\label{eq:DefMusT}
\end{eqnarray}
\endnumparts

Solving $-\partial \ln Z / \partial \eta_i(t)=0$ gives:

\begin{equation}
\fl \hat{\eta}_i(t)=i\mu_i(t+1)-i\frac{\sqrt{\det \bS}}{\left( 2\pi\right)^{NT}}\int \tilde{D}\cG \tanh \left[ g_i(t)+\eta_i(t)\right]\label{eq:SaddleAlphaHat}
\end{equation}

Looking for the stationary points of (\ref{eq:GaussGenFunc1}) with respect to $\bS^{-1}$ corresponds to solving the following set of equations:

\begin{eqnarray}
\fl \frac{\partial \ln Z}{\partial S^{-1}_{ij}(t,t')}=& -\frac{1}{2}S_{ji}(t',t)-i\delta_{ji+N}\delta_{tt'}& \nonumber \\
\fl & +\frac{\det \bS^{1/2}}{\left( 2\pi\right)^{NT}}\sum_{m,s}\int \tilde{D}\cG\left\lbrace \frac{1}{2}\partial _{it,jt'} \ln 2 \cosh \left[ g_m(s)+\eta_m(s)\right] \right\rbrace & \label{eq:saddleS}\\
\fl & -\frac{\det \bS^{1/2}}{\left( 2\pi\right)^{NT}}\sum_{m,s}\int \tilde{D}\cG  \left\lbrace \frac{1}{2}\partial _{it,jt'} \ln 2 \cosh \left[ g_m(s)+\eta_m(s)\right.\right.\nonumber\\
\fl & \left.\left.+i\psi_m(s+1)-i\sum_l J_{lm}\left( \hat{g}_l(s+1)+\hat{\eta}_l(s+1)\right) \right] \right\rbrace & =  0\nonumber
\end{eqnarray}
where we have defined $\partial _{it,jt'}\equiv \frac{\partial^2}{\partial \cG_i(t) \partial \cG_j(t')}$.

\subsection{Equations for the magnetizations}
\label{subsec:ExactSolution}
In the previous subsection we derived expressions for the parameters of the gaussian used for perturbative approximation of the log-generating functional at fixed $\psi$. Now we want to derive an expression for the magnetizations using (\ref{eq:magn0}). We will first perform the derivative of (\ref{eq:lnZTOT}) with respect to $\psi$; notice that even $\be$, $\hat{\be}$ and $\bS$ are $\psi$ dependent, such that (\ref{eq:magn0}) reads:

\begin{eqnarray}
 m_i(t)&=&-i\lim_{\bpsi \to 0} \frac{\partial \ln Z}{\partial \psi_i(t)}+\sum_{j,t'}\frac{\partial \eta_j(t')}{\partial \psi_i(t)}\frac{\partial \ln Z}{\partial \eta_j(t')}+\sum_{j,t'}\frac{\partial \hat{\eta}_j(t')}{\partial \psi_i(t)}\frac{\partial \ln Z}{\partial \hat{\eta}_j(t')}\nonumber\\
&&+\sum_{l,j,t',t''}\frac{\partial S_{l,j}(t',t'')}{\partial \psi_i(t)}\frac{\partial \ln Z}{\partial S_{l,j}(t',t'')}
\label{eq:Fullmagn}
\end{eqnarray}

However, since in our optimization scheme we looked for the stationary values of $\ln Z$ with respect to the variational parameters, $\partial\ln Z / \partial\psi$ will only consist of its explicit derivative with respect to $\psi$, leading to: 

\begin{equation}
m_i(t)=\lim_{\bpsi \to 0}\mu_i(t)
\label{eq:MuPsi}
\end{equation}
for all $t=0,...,T$ and $\mu_i(t)$ has been defined in (\ref{eq:DefMus}), (\ref{eq:DefMus0}) and (\ref{eq:DefMusT}).

\subsection{The optimized values of the parameters}

In principle, one needs to solve the full set of equations (\ref{eq:SaddleAlpha})-(\ref{eq:saddleS}) and take the limit of $\bpsi \to 0$ to calculate the magnetization in \eref{eq:MuPsi}. This is obviously a very difficult task to do analytically given the high dimensional integrals that appear in (\ref{eq:DefMus})-(\ref{eq:saddleS}) and that the equations have to be solved simultaneously. The solutions, however, can be very much simplified if we assume

\begin{equation}
\lim_{\bpsi \to 0} \hat{\eta}_i(t) = 0
\label{eq:HatAlphaPsi0}
\end{equation}
$\forall i, \forall t$.
With (\ref{eq:HatAlphaPsi0}), which we will justify in sec. \ref{subsec:Consistency} below, the optimal interaction matrix $\bS$ in (\ref{eq:saddleS}) in the limit $\bpsi\to 0$ assumes the following block tridiagonal structure:
\begin{equation}
\fl
\bS=
\left[
\begin{array}{cccccc}
\bS(0,0) & \bS(0,1)& 0 &0&0&\cdots\\
\bS(1,0) & \bS(1,1)& \bS(1,2) &0&0&\cdots\\
0&\bS(2,1) & \bS(2,2)& \bS(2,3) &0&\cdots\\
0 &0 & \bS(3,2)& \bS(3,3) &\bS(3,4)&\cdots\\
\cdots&\cdots&\cdots&\cdots&\cdots&\cdots
\end{array}
\right]
\label{eq:matrixS}
\end{equation}
where
\begin{equation}
\bS(t,t)=
\left[
\begin{array}{c|c}
0 & -i\mathbb{I} \\
\hline
-i\mathbb{I}  & \bgam (t) \\
\end{array}
\right]
,
\bS(t,t+1)=
\left[
\begin{array}{c|c}
0 & \blam (t,t+1) \\
\hline
0 & 0 \\
\end{array}
\right],
\label{eq:matrixS1}
\end{equation}
the blocks $\bS(t,t')$ are of size $2N\times 2N$, $\bS(t,t+1)=\bS(t+1,t)^\top$ and 

\numparts
\begin{eqnarray}
\gamma_{ij}(t)&=\sum_k J_{ik}J_{jk}\left( 1-m_k(t)^2\right)\qquad &t=0,...,T-1 \label{eq:gamma}\\
\lambda_{ij}(t,t+1)&=i J_{ji}\left( 1-m_i(t+1)^2\right)\qquad &t=0,...,T-2\label{eq:lambda}
\end{eqnarray}
\endnumparts

Observe that the matrix $\bS$ in (\ref{eq:matrixS}) is a symmetric complex matrix (not hermitian), whose hermitian part is positive symmetric.(Recall that the hermitian part of a matrix $\bS$ is defined as $ (\bS + \bS^\dagger)/2$.) This is consistent with its derivation given that --- as pointed out in \cite{altland2010condensed}--- the gaussian integral $\int \tilde{D}\cG$ converges only if the hermitian part of $ \bS $ is a positive symmetric matrix. 

In (\ref{eq:gamma}) and (\ref{eq:lambda}) we implicitly state that $\det \bS=1$: as a matter of fact it can be proven to be a mere consequence of the block structure of the matrix $\bS$, as shown in \ref{sec:DetS}. Since $\int \tilde{D} \cG = (2 \pi)^{NT}\sqrt{\det \bS}$ this means that the gaussian integral and the model log generating functional match in the limit $\bpsi \to \mathbf{0}$.

Finally we can substitute the optimal values of the variational parameters in (\ref{eq:MuPsi}) and exploit (\ref{eq:SaddleAlphaHat}) to get:

\begin{eqnarray}
\fl m_i(t)= \frac{1}{\left( 2\pi\right)^{NT}}\int \tilde{D}\cG  \  \tanh  [  g_i(t-1) +h_i(t-1)+\sum_j J_{ij}m_j(t-1)] \label{eq:magn01}
\end{eqnarray}
for $t=1,...,T-1$. 

We are now left to evaluate a multidimensional integral in \eref{eq:magn01}.  In fact the integration in \eref{eq:magn01} can be reduced to a one-dimensional integral marginalizing the multivariate gaussian distribution, yielding
\begin{eqnarray}
&& m_i(t)=\int \frac{dx e^{-x^2/2}}{\sqrt{2\pi}} \tanh\left[ x \sigma(t) +h_i(t-1)+\sum_j J_{ij}m_j(t-1)\right]
\label{eq:magn02}\\
&&\sigma(t) = \sqrt{\left( \bS^{-1}\right)_{2N(t-1)+i,2N(t-1)+i}},
\end{eqnarray}
where the integral is now over $x=g_i(t)/ \sqrt{\left( \bS^{-1}\right)_{2N(t-1)+i,2N(t-1)+i}} $, a normally distributed, zero mean unit variance, random variable.

For performing the one dimensional integral in \eref{eq:magn02}, we need to compute the entries of the inverse of matrix $\bS$. In \ref{sec:InvS} we demonstrate that, given $\bS$ as defined in \eref{eq:matrixS}-\eref{eq:lambda}, the entries of $\bS^{-1}$ in which we are interested in can be calculated recursively as

\begin{equation}
\fl \bS^{-1}_{2Nt+i,2Nt+i}=\tilde{\bgam}_{ii}(t)\quad {\rm where}\quad \tilde{\bgam}(t)=\bgam(t)-\blam(t-1,t)^{\top}\tilde{\bgam}(t-1)\blam(t-1,t).
\label{eq:invSrec}
\end{equation}

As we show in \ref{sec:InvS}, $\tilde{\bgam}_{ii}(t)$ can only take positive values and therefore the integral in \eref{eq:magn02} is physically well-defined. 

Recalling the definitions of the matrices $\bgam$ and $\blam$, one can verify that the magnetizations in (\ref{eq:magn02}) only depend on the past magnetizations $m_j(t')$ with $t'<t$, $j=1,\dots,N$. Since this dependence goes back to $t'=0$ it is natural to wonder if the error in estimating the past magnetizations would accumulate impairing the inference process. We notice (not included in section \ref{sec:NumResults}) that for the Gaussian Average method knowledge of the history of the experimental magnetization --- knowing $\tilde{\bgam}_{ii}(t-1)$ when computing $m_i(t)$ with (\ref{eq:magn02}) --- doesn't affects the reconstruction significantly. Whether we are using experimental magnetizations or approximate ones in (\ref{eq:invSrec}), we observe that $\tilde{\bgam}_{ii}(t-1)$ grows exponentially with time for strong couplings while it converges to a finite value for weak couplings. This behavior can be understood by studying the stability of the map (\ref{eq:invSrec}) of $\tilde{\bgamma}(t-1)$ into $\tilde{\bgamma}(t)$ that defines a dynamical system, as we do in \ref{sec:Variance}. Averaging over the disorder one realizes that this dynamical system is chaotic for couplings strength above a certain critical value. Its critical value depends on the degree of symmetry of the connectivity and on the presence of an external field. 

\subsection{The solution $\lim_{\bpsi \to 0} \hat{\eta}_i(t) = 0$}
\label{subsec:Consistency}

In principle, the value of limit of $\bpsi\to 0$ of $\hat{\eta}_i(t)$ that satisfy the optimality equations, may be non-zero. In this section, we justify the choice of $\lim_{\bpsi\to 0}\hat{\eta}_i(t)=0$ that we made in the previous section. We first note that zero is a good candidate for the optimal value of $\hat{\eta}_i(t)=\left\langle \hat{g}_i(t)\right\rangle_{L_s}$ --- here $\left\langle \cdot\right\rangle_{L_s}$ indicates the average under the complex measure $e^{-L_s}$ in (\ref{eq:GaussMeasure})-(\ref{eq:GaussForm}) --- since

\begin{equation}
\lim_{\bpsi \to 0} \left\langle \hat{g}_i(t)\right\rangle_{L}=0,
\end{equation} 
where $L$ for the kinetic Ising model has been defined in (\ref{eq:GenFuncIsing}) and $\left\langle \cdot\right\rangle_{L}$ indicates the average under the complex measure $e^{-L}$ . This choice for the mean in the $\hat{g}$s can be justified by analogy with the mean in the $g$s: the stationary value for latter is also the saddle point value of the kinetic Ising generating functional, while the saddle point in the $\hat{g}$s is conventionally set to zero. 

Furthermore, we can show that $\lim_{\bpsi \to 0} \hat{\eta}_i(t) = 0$ yields a consistent solution. To do this we first note that by inverting the matrix $\bS$, as shown in \ref{sec:InvS}, two point correlation functions $\left\langle g_i(t-1)\hat{g}_j(t)\right\rangle_{L'_s}$ and $\left\langle \hat{g}_i(t)\hat{g}_j(t)\right\rangle_{L'_s}$ are both zero, where notation $\left\langle \cdot\right\rangle_{L'_s}$ indicates averages under the gaussian measure $e^{-L'_s}$, with $L'_s=\frac{1}{2}\cG \bS \cG$. 
Consequently, we have

\begin{eqnarray}
\fl \lim_{\bpsi \to 0}\mu_i(t)&=&\left\langle  \tanh \left[ g_i(t-1)+\eta_i(t-1)-i\sum_l J_{li}\hat{g}_l(t)\right]\right\rangle_{L'_s} \nonumber\\
\fl &=&\sum^{\infty}_{n=1} a_n \left\langle \left[ g_i(t-1)+\eta_i(t-1)-i\sum_l J_{li}\hat{g}_l(t)\right]^{2n-1}\right\rangle_{L'_s} \nonumber\\
\fl &=&\sum_{n,k,l} b_{n,k,l} \left( \eta_i(t-1)\right)^{k-l} \left\langle g_i(t-1)^l\left( -i\sum_l J_{li}\hat{g}_l(t)\right)^{2n-1-k}\right\rangle_{L'_s} \nonumber\\
\fl &=&\sum_{n,k,l} b_{n,k,l} \delta_{k,2n-1}\left( \eta_i(t-1)\right)^{k-l} \left\langle g_i(t-1)^l\right\rangle_{L'_s} \nonumber\\
\fl &=&\left\langle  \tanh \left[ g_i(t-1)+\eta_i(t-1)\right]\right\rangle_{L'_s}\label{eq:HatAlpha0Cons}
\end{eqnarray}
Now, note that the previous equality corresponds to setting $\hat{\eta}_i(t)=0$ in (\ref{eq:SaddleAlphaHat}).

\subsection{The fully asymmetric limit}\label{subsec:VarG_FullyAsy}
In \cite{mezard2011exact} Mezard and Sakellariou derive equations for the magnetizations that are exact for fully asymmetric couplings:
\begin{equation}
\fl m_i(t)=\int \frac{dx}{\sqrt{2\pi}} e^{-x^2/2} \tanh\left[  h_i(t-1) + \sum_j J_{ij}m_j(t-1)+x\sqrt{\gamma_{ii}(t-1)}\right]
\label{eq:ExactMF}
\end{equation}
where $\gamma_{ii}(t)$ has been defined in (\ref{eq:gamma}). 

In section \ref{sec:Optimization} all entries of $\bS$ were free to be optimized. However, we could have assumed that the blocks corresponding to $\blam(t-1,t)$ are set to zero a priori. By looking at (\ref{eq:magn02}) and (\ref{eq:invSrec}) one easily realizes that with this constraint our optimization would have lead to (\ref{eq:ExactMF}), which is exact in the fully asymmetric limit.  Notice that this prescription on $\blam$ would not affect the optimal value of any other variational parameter, since we optimized $\ln Z$ independently with respect of $\be$, $\hat{\be}$ or $\bS_{ij}(t,t')$.

\section{Extended Plefka expansion}
\label{sec:ExtPlefka}
As mentioned in the Introduction, two particularly powerful approaches to studying disordered systems both in machine learning and statistical physics community are variational and weak coupling  expansions. In the previous sections we reported some results regarding the variational approach. In this section we aim at developing a comprehensive weak coupling expansion for the disordered spin systems. 

Weak coupling expansions in field theory and statistical physics of disordered systems take several forms. One of the most powerful amongst these, which has proven to be particularly useful for studying the equilibrium properties of glassy systems, is the Plefka expansion. The Plefka expansion was originally performed for the equilibrium Sherrington-Kirckpatrick model by 
expanding the Gibbs free energy at fixed magnetization, enforced via a Legendre transform, around 
the free energy of an uncoupled system. To the first order in $J$ it yields the naive mean field 
results while to the second order the TAP equations are recovered. Although higher order terms vanish 
for the SK model, they can in general be computed \cite{georges1991expand}.  

In performing the Plefka expansion for the equilibrium model with binary spins it is sufficient to fix the magnetization
and this has been the line taken by Roudi and Hertz in deriving Plefka expansion and dynamical TAP equations for the kinetic Ising model.
However, in contrast to the equilibrium case, for the dynamics the magnetization is not the only relevant order parameter.  
Including other observables in the Plefka expansion, namely the correlation and response functions, is what we do in this section. As we will show with numerical results in the next section, this will lead to a significant improvement for predicting the dynamics of the system.  

Instead of the generating functional in (\ref{eq:GenFunc1}) and (\ref{eq:GenFunc2}), let us now consider the following functional:
\begin{equation} 
\fl Z_{\alpha}[{\boldsymbol \psi}, { \boldsymbol h}, \hat{\bold C},\hat{\bold B},\hat{\bold R}]= \frac{1}{2^N (2 \pi)^{NT}} \int D\mathcal{G}
 \, {\rm Tr}_s \exp \left( L_{\alpha}[{\boldsymbol \psi}, { \boldsymbol h}, \hat{\bold C},\hat{\bold B},\hat{\bold R}, {\mathcal G},{\bold s}] \right) ,
\label{eq:gen_fun}
\end{equation}
with
\begin{eqnarray}
\fl L_{\alpha}[{\boldsymbol \psi}, { \boldsymbol h}, \hat{\bold C},\hat{\bold B},\hat{\bold R}, {\mathcal G},{\bold s}]  = \sum_{i,t}  \left\{ i \hat{g}_i(t)\left[ g_i(t) - \alpha \sum_j J_{ij}s_j(t) \right] + s_i(t+1) g_i(t) 
 \nonumber \right.\\
 \fl \left. - \ln 2 \cosh g_i(t)  -i h_i(t) \hat{g}_i(t)  + \psi_i(t) s_i(t)+ \frac{1}{2} \sum_{t'} \hat{C}_i(t,t') s_i(t) s_i(t')  \right.\nonumber\\
\fl  \left. + \frac{1}{2} \sum_{t'} \hat{B}_i(t,t') \hat{g}_i(t) \hat{g}_i(t')
 -i \sum_{t'}\hat{R}_i(t',t) \hat{g}_i(t) s_i(t')  \right\},
 \label{eq:Plefka_action}
\end{eqnarray}
where we have introduced the parameter $\alpha$ to control the  interaction strength. 
The introduction of the new auxiliary fields $\hat{C}, \hat{B}$ and $\hat{R}$ in the action (\ref{eq:Plefka_action}) is related to the averages of the observables that we want to
constrain when performing the Legendre transform.
In particular, here we decide to fix all marginal first and second moments over time.
One can find the moments and the physical meaning of these auxiliary fields 
by first derivatives of the generating functional with respect to the fields
as follows:
\begin{eqnarray}
-i\hat{m}_i(t) &=& \frac{\partial \ln Z_{\alpha}}{\partial h_i(t)}=-i \langle \hat{g}_i(t) \rangle_{\alpha}\nonumber\\
 m_i(t) &=& \frac{\partial \ln Z_{\alpha}}{\partial \psi_i(t)}= \langle s_i(t) \rangle_{\alpha}\nonumber\\
 C_i(t,t') &=& \frac{\partial \ln Z_{\alpha}}{\partial \hat{C}_i(t,t')}=\langle s_i(t) s_i(t') \rangle_{\alpha} \quad {\rm for} \; t'\neq t  \nonumber\\
 \frac{1}{2} B_i(t,t') &=& \frac{\partial \ln Z_{\alpha}}{\partial \hat{B}_i(t,t')}= \frac{1}{2} \langle \hat{g}_i(t) \hat{g}_i(t') \rangle_{\alpha}\nonumber\\
 R_i(t,t') &=& \frac{\partial \ln Z_{\alpha}}{\partial \hat{R}_i(t,t')}= - i \langle \hat{g}_i(t') s_i(t) \rangle_{\alpha}\,,
\label{eq:moments}
\end{eqnarray}
where $\langle \cdots \rangle_{\alpha}$ denotes averaging 
over the distribution defined by the measure inside the functional (\ref{eq:gen_fun}). Namely,
for any function $F(\bold s)$ of the trajectory of spins $\bold s$  we define:
\begin{equation} 
\langle F \rangle_{\alpha} = \frac{{ \int D\mathcal{G}
 \, {\rm Tr}_s \; F(\bold s) \; \exp \left( L_{\alpha}[{\boldsymbol \psi}, { \boldsymbol h}, \hat{\bold C},\hat{\bold B},\hat{\bold R}, {\mathcal G},{\bold s}] \right)} }{ \int D\mathcal{G}
 \, {\rm Tr}_s \exp \left( L_{\alpha}[{\boldsymbol \psi}, { \boldsymbol h}, \hat{\bold C},\hat{\bold B},\hat{\bold R}, {\mathcal G},{\bold s}] \right)} .
 \label{eq:average_alpha_def}
\end{equation}
The moments of the original dynamical system (\ref{eq:PModel0})
can be found by setting the auxiliary fields to zero and $\alpha=1$ at the end of the calculation.
Note that ${C}(t,t)=1$ and its conjugate field $\hat{C}(t,t)=0$.

The Legendre transform of $\ln Z$ 
is given by 
\begin{eqnarray}
\fl \Gamma_{\alpha} [{\bold m},{\hat{\bold m}}, {\bold C},  {\bold B},{\bold R}]=\ln Z_{\alpha}[{\boldsymbol \psi}, { \boldsymbol h}, \hat{\bold C},\hat{\bold B},\hat{\bold R}]
- \sum_{it}\psi_i(t)m_i(t) +i\sum_{it}h_i(t)\hat{m}_i(t) \nonumber\\
\fl-\frac{1}{2} \sum_{itt'} \hat{C}_i(t,t') C_i(t,t')  - \frac{1}{2} \sum_{itt'} \hat{B}_i(t,t') B_i(t,t') 
 - \sum_{itt'} \hat{R}_i(t',t) R_i(t',t), 
\label{eq:Gamma_def}
\end{eqnarray}
where the fields
${\boldsymbol \psi}, { \boldsymbol h}, \hat{\bold C},\hat{\bold B},\hat{\bold R} $
in the above equation are to be considered as functions
of the moments, and dependent on the $\alpha$ parameter, 
according to the following set of equations:
\begin{eqnarray}
 \frac{\partial \Gamma_{\alpha}}{\partial m_i(t)}&=&-\psi_i^{\alpha}[{\bold m},{\hat{\bold m}}, {\bold C},  {\bold B},{\bold R}](t)\nonumber\\
 \frac{\partial \Gamma_{\alpha}}{\partial \hat{m}_i(t)}&=&ih_i^{\alpha}[{\bold m},{\hat{\bold m}}, {\bold C},  {\bold B},{\bold R}] (t)\nonumber\\
 \frac{\partial \Gamma_{\alpha}}{\partial C_i(t,t')}&=&-\hat{C}_i^{\alpha}[{\bold m},{\hat{\bold m}}, {\bold C},  {\bold B},{\bold R}](t,t')\quad {\rm for} \; t'\neq t \nonumber \\
 \frac{\partial \Gamma_{\alpha}}{\partial B_i(t,t')}&=&-\frac{1}{2} \hat{B}_i^{\alpha}[{\bold m},{\hat{\bold m}}, {\bold C},  {\bold B},{\bold R}](t,t')\nonumber\\
 \frac{\partial \Gamma_{\alpha}}{\partial R_i(t,t')}&=&- \hat{R}_i^{\alpha}[{\bold m},{\hat{\bold m}}, {\bold C},  {\bold B},{\bold R}](t,t'). 
\label{eq:param}
\end{eqnarray}
We now perform a second order expansion of $\Gamma_{\alpha}$ around $\alpha=0$ 
and consider the set of equations (\ref{eq:param}) within the expansion; the details of the calculation are reported in \ref{sec:Plefka}.
Setting the auxiliary fields to zero, we can extract the value of the fields
${\boldsymbol \psi}^0, { \boldsymbol h}^0, \hat{\bold C}^0,\hat{\bold B}^0,\hat{\bold R}^0$
as functions of the correct (within the expansion) marginal first and second moments. 
Those fields thus represent effective external fields 
which have to 
be applied to the model {\em without} interactions ($\alpha =0$) to obtain the same
 moments as the interacting model.
 Hence, we may consider
$Z_0[{\boldsymbol \psi}^0, { \boldsymbol h}^0, \hat{\bold C}^0,\hat{\bold B}^0,\hat{\bold R}^0]$
 as the generating functional for the true marginal distributions, giving us
an effective noninteracting description of the true
interacting dynamics. 
The explicit calculation (\ref{sec:Plefka}) yields $Z^0[h]=\prod_i Z^0_i[h_i]$, where

\begin{eqnarray}
\fl & Z^0_i \propto \left\langle \int d\textbf{g}_i  {\rm Tr}_{s_i}   \prod_{t}
\frac{e^{s_i(t+1) g_i(t)}}{2 \cosh g_i(t)} \prod_{t}  \delta \left[ \vphantom{ \left( \sum_{t=0}^{k-2} \right) } g_i(t)   \right.  \right. \\
 \fl & \left. \left. -\phi_i(t)  - \sum_j \left( J_{ij} m_j(t) 
 -  \sum_{t'=0}^{t-1} J_{ij}J_{ji} R_j(t,t') [s_i(t') - m_i(t') ]\right) - h_i(t)  \right] 
\right\rangle_{\phi_i}  \nonumber
\label{eq:fin_Z}
\end{eqnarray}
and where $\phi_i(t)$ is a Gaussian random variables, drawn independetly for each $i$, with zero mean and 
covariance 
\begin{equation}
\langle \phi_i(t) \phi_i(t') \rangle = \sum_j J^2_{ij} [C_j(t,t') - m_j(t)m_j(t')].
\label{eq:noise_covariance}
\end{equation}
This corresponds to a stochastic equation for a single spin, where each spin $i$ is subjected to an effective field 
\begin{equation} 
\fl g_i(t)  = \phi_i(t)  + \sum_j \left( J_{ij} m_j(t) 
 - \sum_{t'=0}^{t-1} J_{ij}J_{ji} R_j(t,t') [s_i(t') - m_i(t') ]\right) + h_i(t).
\label{eq:h}
\end{equation}
The effective field in (\ref{eq:h}) is composed of a coloured Gaussian noise ($\phi$), a naive mean field (the second term), a retarded interaction with the past values of the spins (third term) and finally the external field ($h_i(t)$).

 The retarded interactions and the noise covariance have to be computed 
 as averages from the entire ensemble of independent spins. Luckily, this can be done 
 in a causal fashion, i.e. the spin dynamics depends only on {\em past} spin history.
 However, this can not be done analytically, although one may proceed again with a perturbation 
 expansions in order to get equation of motions for one and two time functions. The fact that the external noise
 is Gaussian should be helpful.  As an alternative, we have resorted to numerical simulations, where the
 necessary averages are estimated from a large number $N_T$ of samples of trajectories.
Sample averages will be denoted by overbars; namely,  for any function $F^k ({\bold s^k})$ of the $k^{\rm th}$ trajectory of spins ${\bold s^k}$ 
we define the following average:
 \begin{equation} 
\overline{ F } = \frac{1}{N_T} \sum_{k=1}^{N_T} F^k.  
 \end{equation} 
In order to compute the retarded interaction $R_i(t,t')$, we recall that 
given a vector ${\boldsymbol \phi}$ with Gaussian distributed components  $\phi(t)$, 
with zero mean and covariance matrix
 $\langle \phi(t) \phi(t') \rangle ={\mathcal{C}}(t,t')$, and given a  
function
$F({\boldsymbol \phi})$ of the vector ${\boldsymbol \phi}$,
the following relation holds
\begin{equation} 
\langle  F({\boldsymbol \phi}) \phi(t) \rangle= \sum_{t'} {\mathcal{C}}(t,t') \left\langle  \frac{\partial}{\partial \phi(t')} F({\boldsymbol \phi}) \right\rangle,
\end{equation}
as can be shown using integration by parts.
By considering the function $F({\boldsymbol \phi})=s(t) \equiv s(t; {\boldsymbol \phi})$ 
and using (\ref{eq:respons}) one finds the following equation relating 
the response and correlation functions:
\begin{equation} 
\langle   s_i(t) \phi_i(t') \rangle_{\phi_i} = \sum_{\tau=1}^{t-1} R_i(t,\tau) \sum_j J^2_{ij} 
[ C_j(\tau, t') -m_j(\tau)m_j(t') ].
\label{eq:resp_plefka}
\end{equation}
The algorithm can be described as follows.  
\begin{itemize}
\item Initial condition: set $s_i^k(0)=1, \quad i=1...N, \, k=1...N_T.$
\item For $t=1...T$:
\begin{enumerate}
\item Draw the spins at time $t$ from
$$
p(s_i^k(t))= \frac{e^{s_i^k(t) g_i^k(t-1)}}{2 \cosh g_i^k(t-1)},\quad {\rm for}\,i=1...N, \, k=1...N_T,
$$
using the fields $g_i^k(t-1)$ calculated at the previous time step.
\item Compute the sample averages
$$
C_i(t,t')= \overline{\tanh[g_i(t-1)] s_i(t')},\quad {\rm for} \,t'=1...t-1,\, i=1...N, 
$$
\item Draw the noise variables 
$
\phi_i^k(t)\: {\rm for} \, i=1...N, \, k=1...N_T,
$
from the conditional probability
$
p(\phi_i^k(t) \vert \phi_i^k(0)...\phi_i^k(t-1)),
$
which can be computed using the Yule Walker equations (\ref{sec:Walker}).
\item Compute the sample averages that will be needed in (\ref{point_v}):
$$
\overline{s_i(t) \phi_i(t')}= \overline{\tanh[g_i(t-1)] \phi_i(t')},\quad {\rm for} \, t'=1...t-1,\, i=1...N.
$$
\item \label{point_v}Compute $R_i(t,t'),\: {\rm for} \, t'=1...t-1$ using 
(\ref{eq:resp_plefka})
by solving the system of linear equations:
$$
\overline{s_i(t) \phi_i(t')}= \sum_{\tau=1}^{t-1} R_i(t,\tau) \sum_j J^2_{ij} [ C_j(\tau, t') -m_j(\tau)m_j(t')].
$$
\item Compute the fields
\begin{eqnarray*}
\fl g_i^k(t)=\phi^k_i(t)  + \sum_j \left( J_{ij} m_j(t) 
 - \sum_{t'=0}^{t-1} J_{ij}J_{ji} R_j(t,t') [s^k_i(t') - m_i(t') ]\right) + h_i(t) , \\ 
\fl {\rm for} \, i=1...N, \, k=1...N_T.
\end{eqnarray*}
\item Compute the magnetizations at time $t+1$:
$$
m_i(t+1)=\overline{\tanh[g_i(t)]} ,\quad {\rm for} \, i=1...N.
$$
\end{enumerate} 
\end{itemize}
To conclude this section, let us point out that
the mean field result (\ref{eq:ExactMF}), which is exact for asymmetric networks in the thermodynamic limit
for Gaussian couplings with variance $1/N$,  can be obtained in two ways.  
One either considers the result (\ref{eq:fin_Z}) and neglects the term $J_{ij}J_{ji}$ for an asymmetric network
 in the limit of  large $N$,
or one works with a simplified Plefka expansion where all two-time moments for different times are excluded from the 
beginning. Hence, from the second moments, one keeps only $B(t,t)$ in the expansion.

\section{Numerical results}
\label{sec:NumResults}
In the previous sections we studied analytically two approaches to improve on the saddle point approximation to the generating functional of the kinetic Ising model with synchronous update. In section \ref{subsec:VarG_FullyAsy} we have argued that the constrained Gaussian Average optimization leads to the Mean Field (MS-MF) equations of \cite{mezard2011exact}, whose performances was studied in \cite{sakellariou2012effect}. One could wonder how this compares to the unconstrained Gaussian Average method, and so we iterated (\ref{eq:magn02}) and (\ref{eq:invSrec}) to reconstruct the entire dynamics of magnetizations.  In order to estimate the magnetizations for the Extended Plefka expansion described in the previous section we designed the algorithm explained in section \ref{sec:ExtPlefka}. Thus we can evaluate numerically the goodness of the two approximations in terms of magnetizations and compare them with existing algorithms. Specifically we investigate how they perform with respect to three mean field methods, namely Naive Mean Field, dynamical TAP (RH-TAP) equations of \cite{roudi2011dynamical} and MS-MF equations of \cite{mezard2011exact}. To recapitulate, Na\"{i}ve Mean Field and TAP equations can be obtained via perturbative expansion in the magnitude of the couplings of the Legendre transform of the log generating functional at fixed magnetizations \cite{roudi2011dynamical}, without making any restriction on symmetry and distribution of the couplings. The first order expansion gives Na\"{i}ve Mean Field, while second order terms lead to RH-TAP. MS-MF equations can be derived via central limit theorem arguments exploiting the fact that the couplings are independent identically distributed  random variables with variance that scales as $1/N$ \cite{mezard2011exact}, without making any assumption on the couplings strength. 

RH-TAP magnetizations under the kinetic Ising model with synchronous update are:

\begin{equation}
m_i(t)= \tanh\left[ h_i(t-1) + \sum_j J_{ij}m_j(t-1)- m_i(t)\gamma_{ii}(t-1) \right]   
\end{equation}
where $\gamma_{ii}(t-1)$ has been defined in (\ref{eq:gamma}). MS-MF equations correspond to (\ref{eq:ExactMF}) and Na\"{i}ve Mean Field to (\ref{eq:NMF}).

In order to test the performances of our methods as a function of couplings asymmetry and strength we chose our couplings, following Crisanti and Sompolinsky \cite{crisanti1987dynamics}:

\begin{equation}
J_{ij}=J_{ij}^{\rm sym}+k J_{ij}^{\rm antisym}
\label{eq:Couplings1}
\end{equation}  
where $J_{ij}^{\rm sym}=J_{ji}^{\rm sym}$ and $J_{ij}^{\rm antisym}=-J_{ji}^{\rm antisym}$, while $k$ is the parameter that controls the asymmetry, interpolating between the fully asymmetric $(k=1)$ and the fully symmetric $(k=0)$ distributions. We draw all the couplings $J_{ij}^{\rm sym}$ and $J_{ij}^{\rm antisym}$ independently from a  distribution with zero mean and variance:

\begin{equation}
\left\langle \left( J_{ij}^{\rm sym}\right)^2 \right\rangle =\left\langle \left( J_{ij}^{\rm antisym}\right)^2 \right\rangle= \frac{g^2}{N\left( 1+k^2\right) } 
\label{eq:Couplings2}
\end{equation}    
where g controls the strength of the couplings.

We initialize the algorithms with the same initial condition and then we iterate them for reconstructing the whole dynamics of magnetizations. We compare the predicted magnetizations with the experimental ones computing the Mean Square Errors:

\begin{equation}
MSE= \frac{1}{T}\frac{1}{N}\left\langle \sum_{i=1}^N\sum_{t=1}^T \left( m_i(t)-m_i^{\rm exp}(t)\right)^2 \right\rangle_{\mathrm{J \, realizations}}
\label{eq:MSE} 
\end{equation}
where $m_i^{\rm exp}(t)$ are obtained by sampling the kinetic Ising model distribution of (\ref{eq:PIsing0}).

\begin{figure}[htp]

\centering

\includegraphics[width=0.48\linewidth]{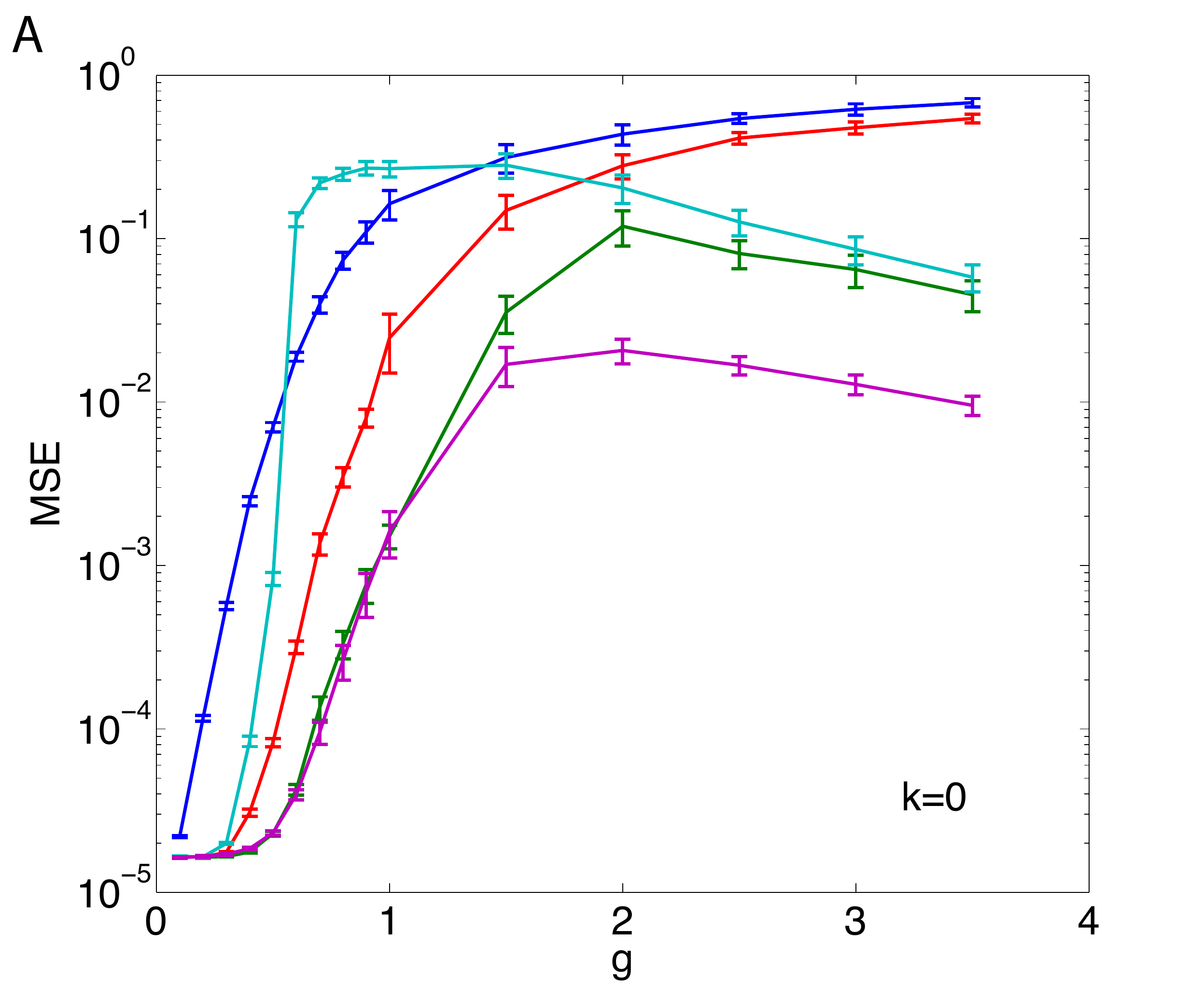} 
\includegraphics[width=0.48\linewidth]{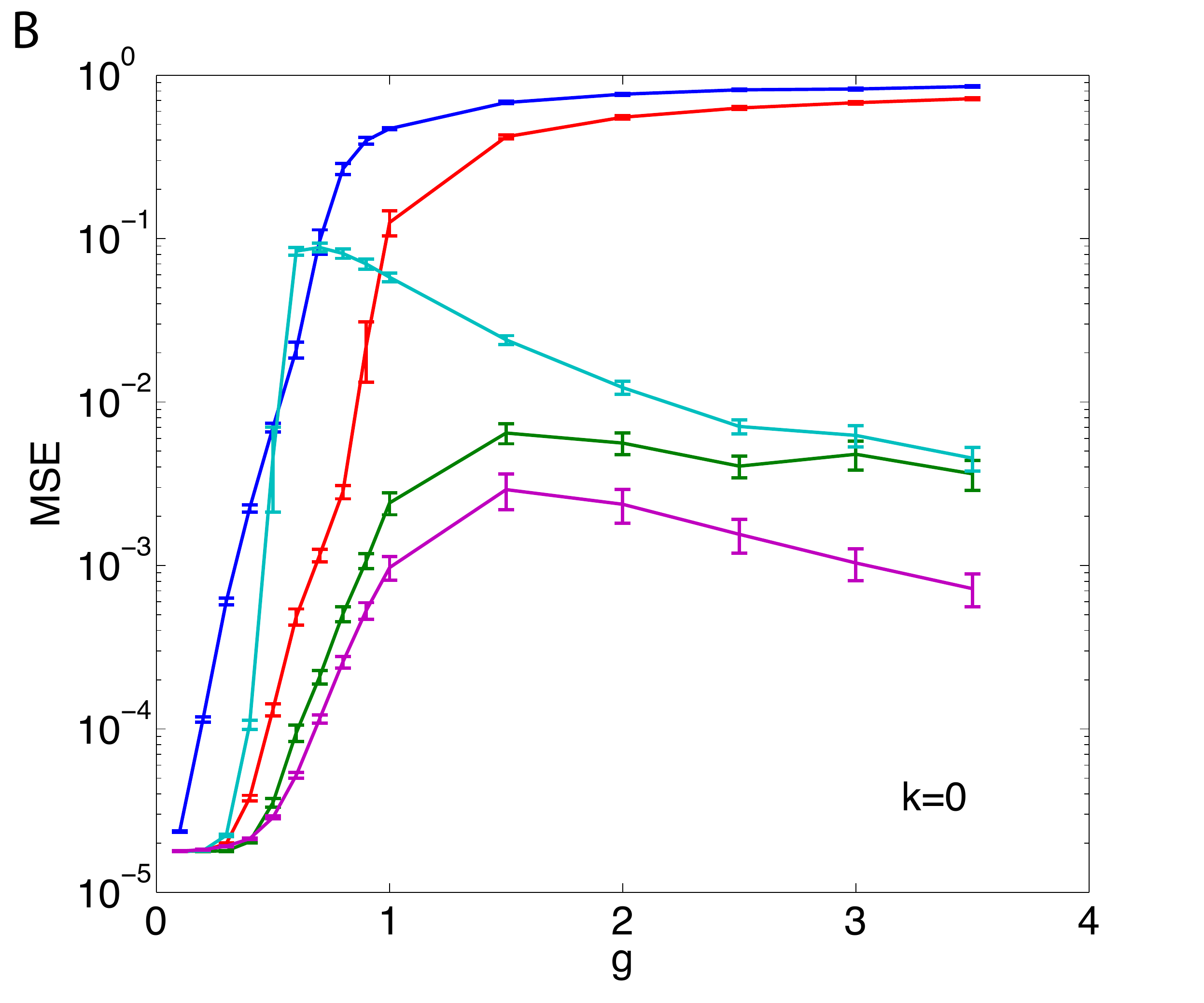}\\

\includegraphics[width=0.48\linewidth]{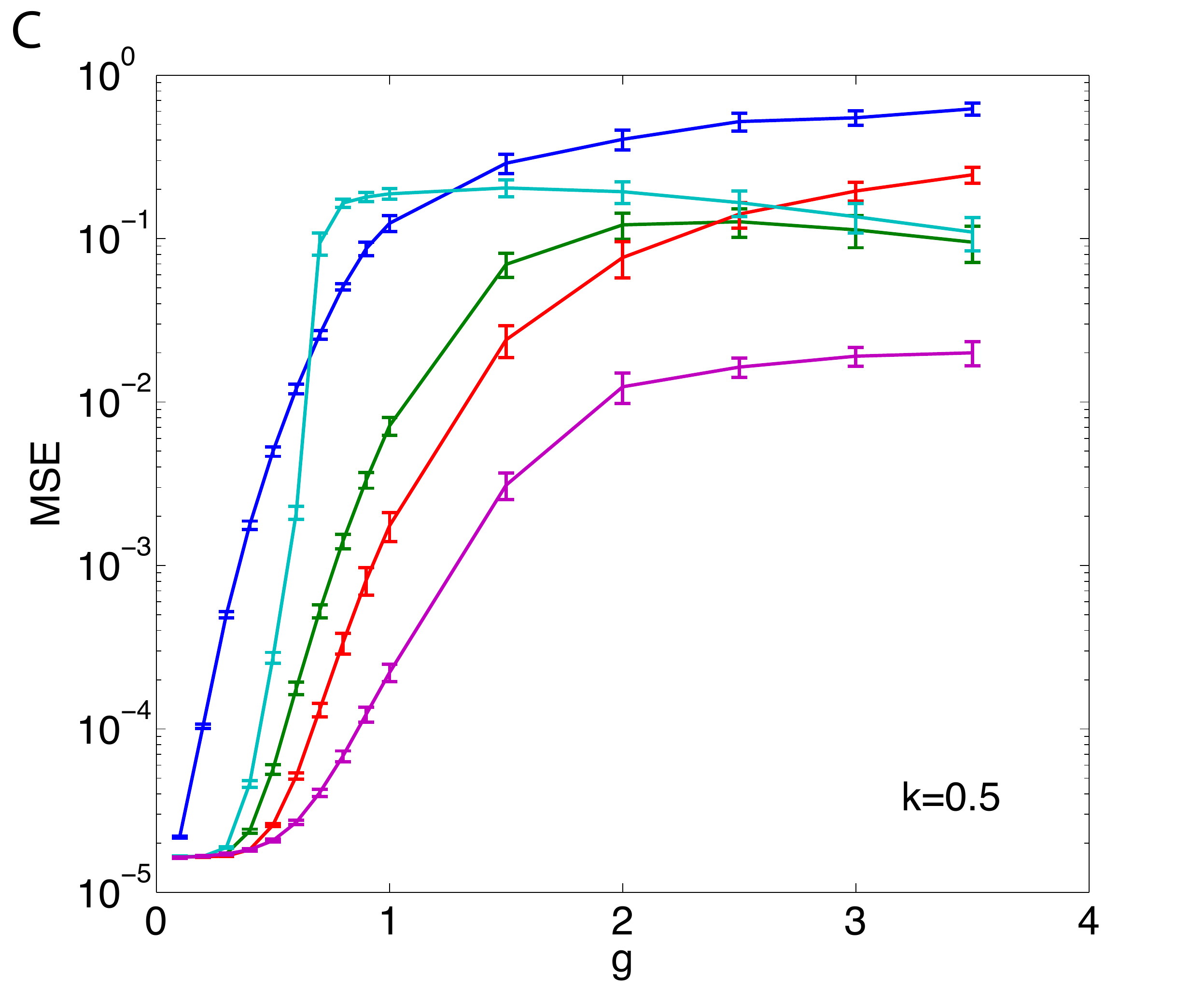} 
\includegraphics[width=0.48\linewidth]{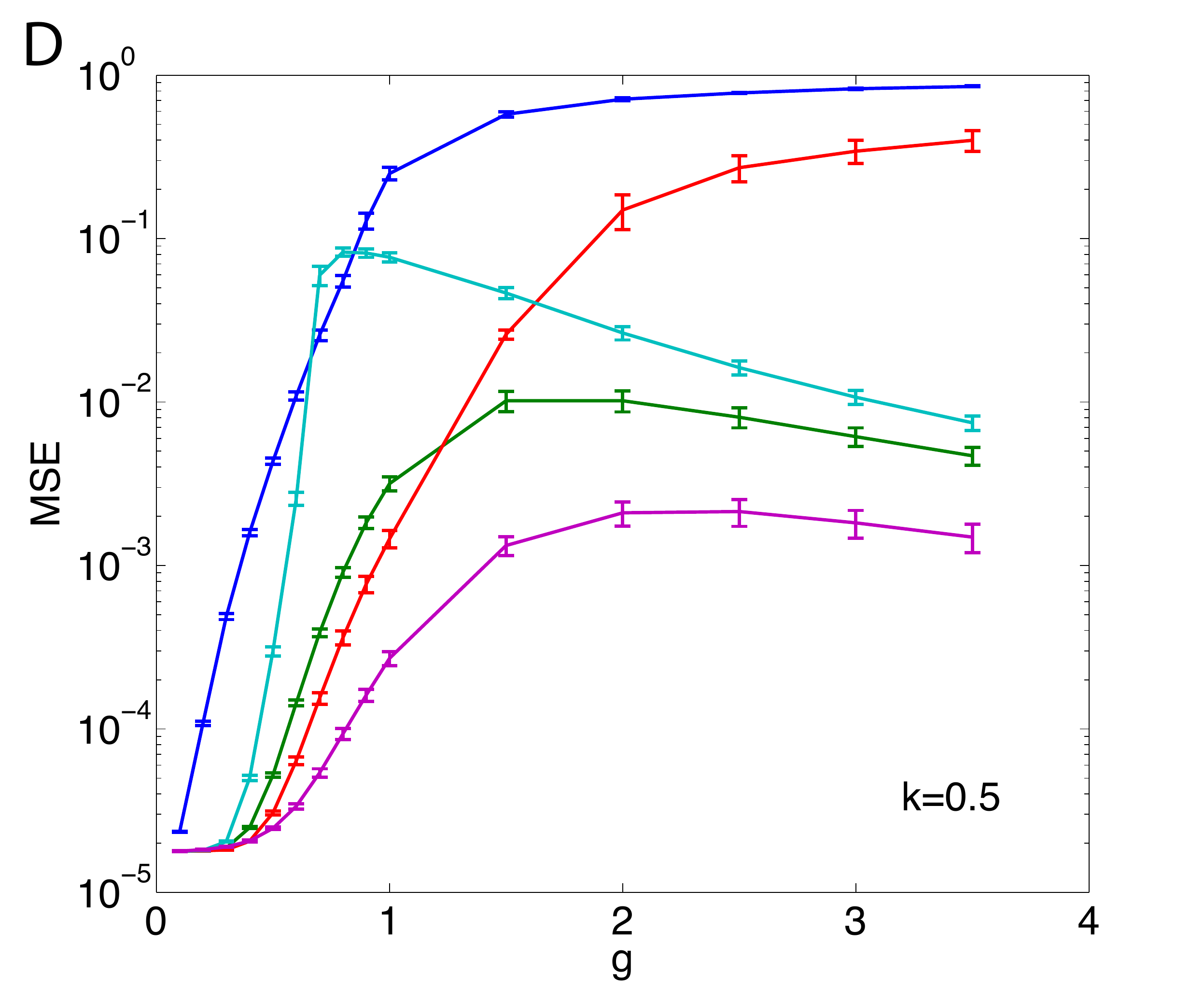}\\

\includegraphics[width=0.48\linewidth]{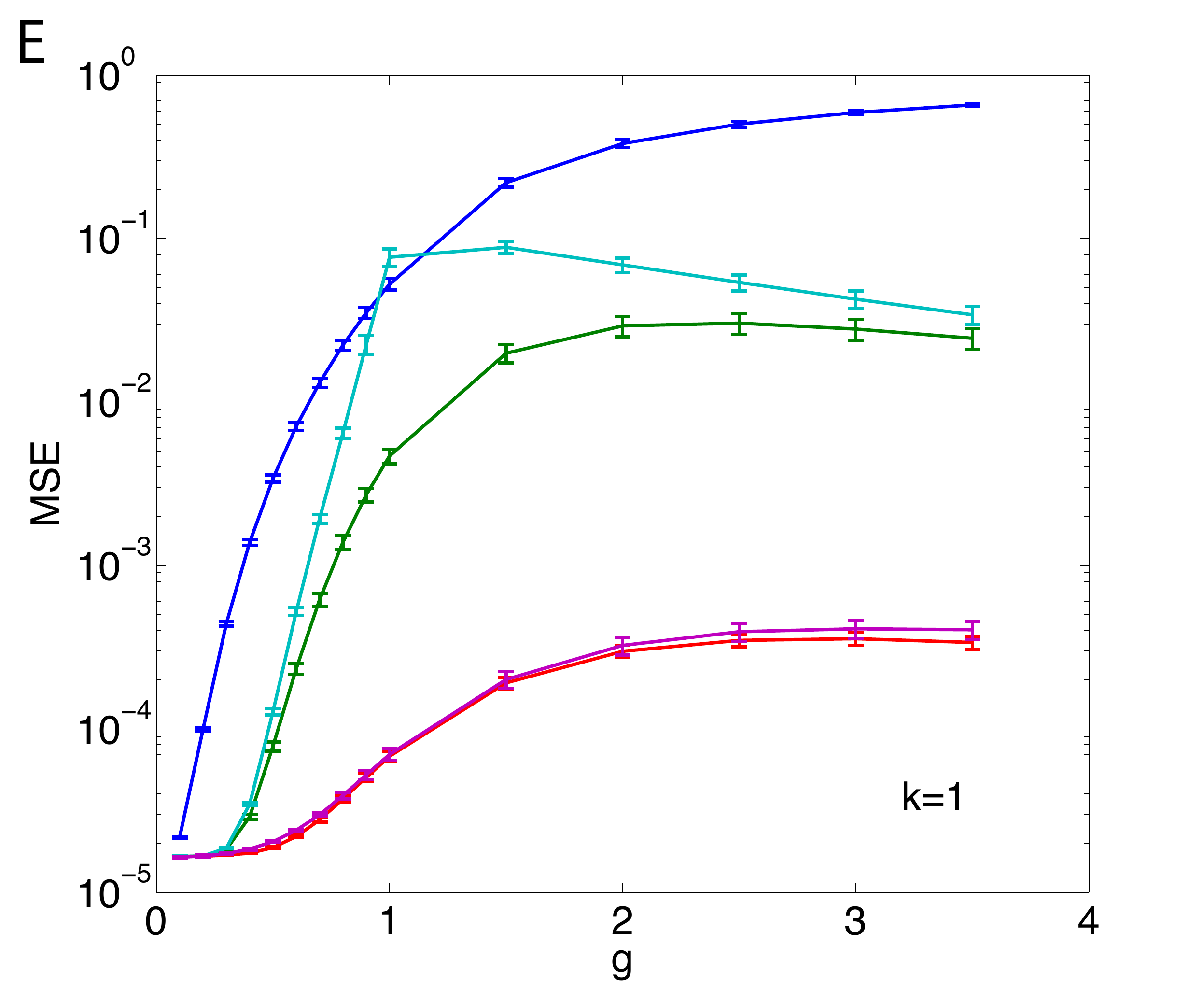}
\includegraphics[width=0.48\linewidth]{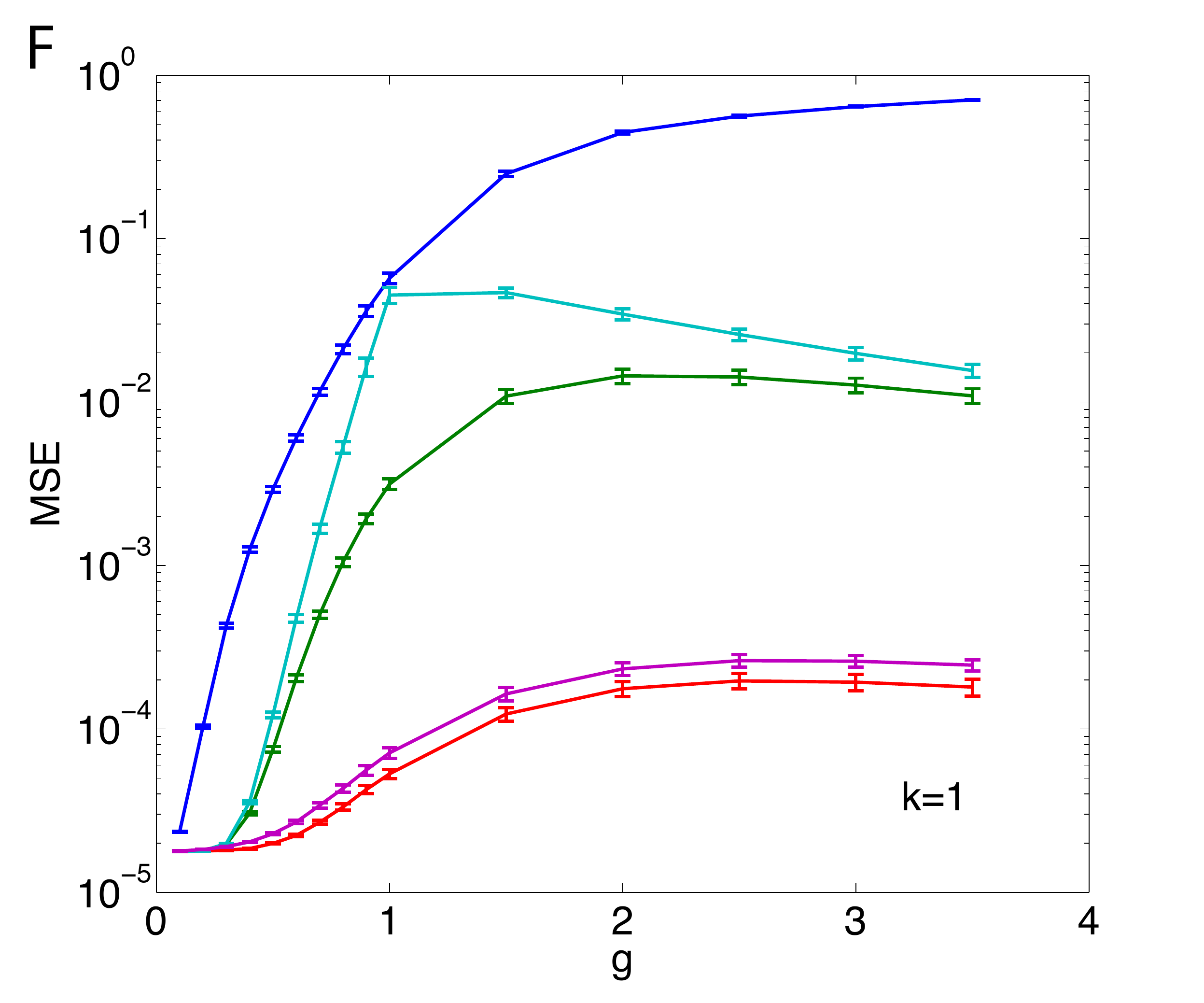}

\caption{Mean squared error of N\"{a}ive Mean field (blue), RH-TAP (green), MS-MF (red), unconstrained Gaussian Average approach (light blue) and Extended Plefka (magenta) for predicting entire dynamics of the magnetizations. The mean squared error is plotted as a function of the couplings strength g for a system of 50 spins. We have used 100 time steps and 50,000 repeats to calculate the experimental magnetizations and have averaged the errors over 10 realizations of the couplings. The error bars are standard deviations over these realizations.The number of sample trajectories used in the algorithm for the Extended Plefka method is $N_T=50000$.
The different panels correspond to different values of the asymmetry parameter $k=0,0.5,1$ from top to bottom. Left: stationary external field drawn independently for each spin from a normal distribution (zero mean, standard deviation 0.5). Right: sinusoidal external field with amplitude $0.5$.  }
\label{fig:StatSinField}
\end{figure}

The results are shown in Fig.\ref{fig:StatSinField}. From these plots it's clear that, apart from Na\"{i}ve Mean field, all methods that we considered are compatible in the high temperature limit. At lower temperatures the Extended Plefka expansion is superior independently of the external field.
Note, however, that for fully asymmetric couplings and sinusoidal external field
the MS-MF method is performing slightly better than the Extended Plefka approximation.
This is likely due to the finite size effects, since 
the two approaches are equivalent for asymmetric networks with large $N$, as explained in section \ref{sec:ExtPlefka}. 
Regardless the degree of symmetry of the couplings and the external field RH-TAP systematically
 improves on the unconstrained Gaussian Average approach, which  fails at intermediate temperatures. 
The fact that the reconstruction is noisier with respect to \cite{sakellariou2012effect} is due to error propagation during the dynamics.

\begin{figure}[htp]

\centering

\includegraphics[width=0.495\linewidth]{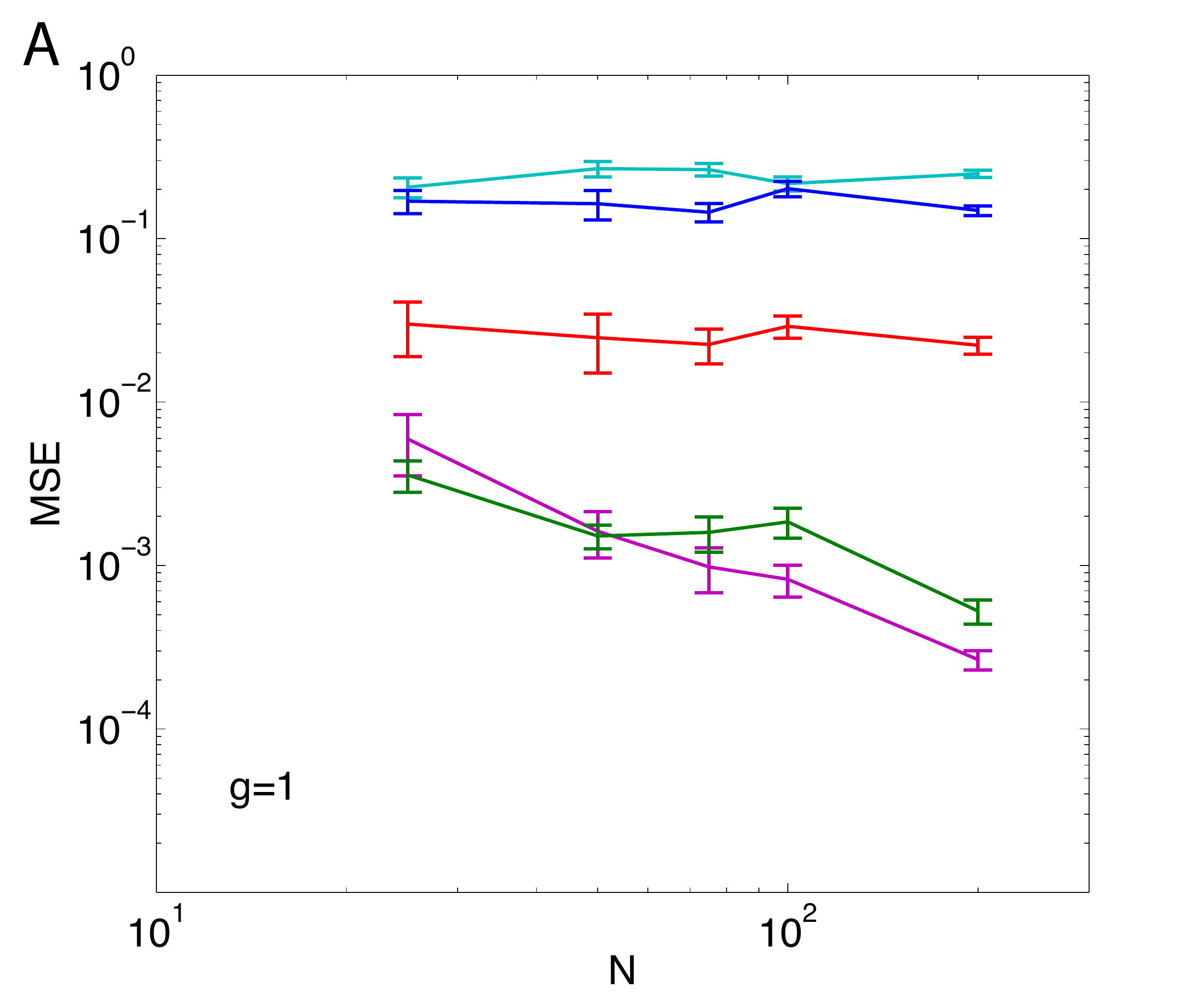} 
\includegraphics[width=0.495\linewidth]{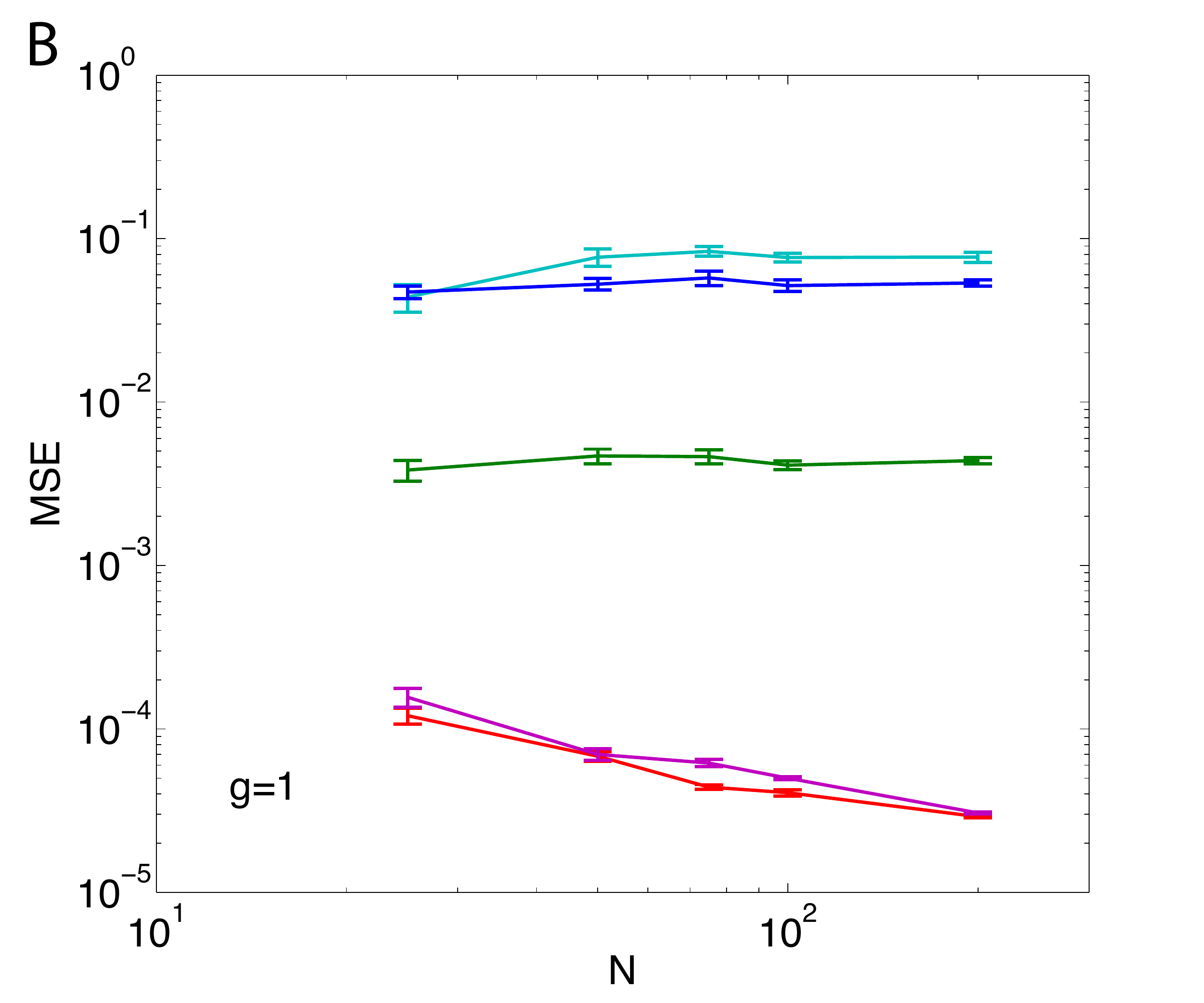}\\
\includegraphics[width=0.495\linewidth]{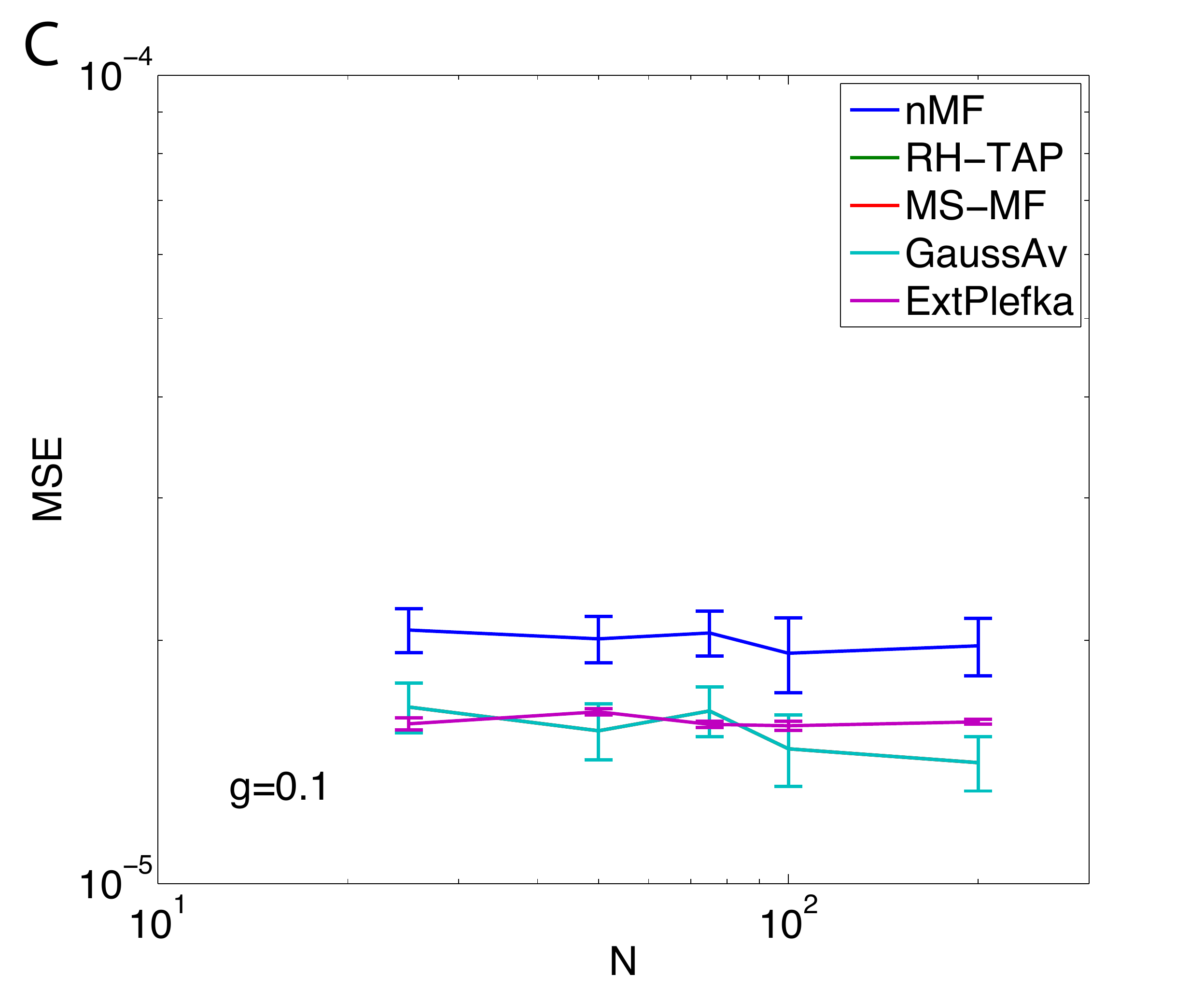} 
\includegraphics[width=0.495\linewidth]{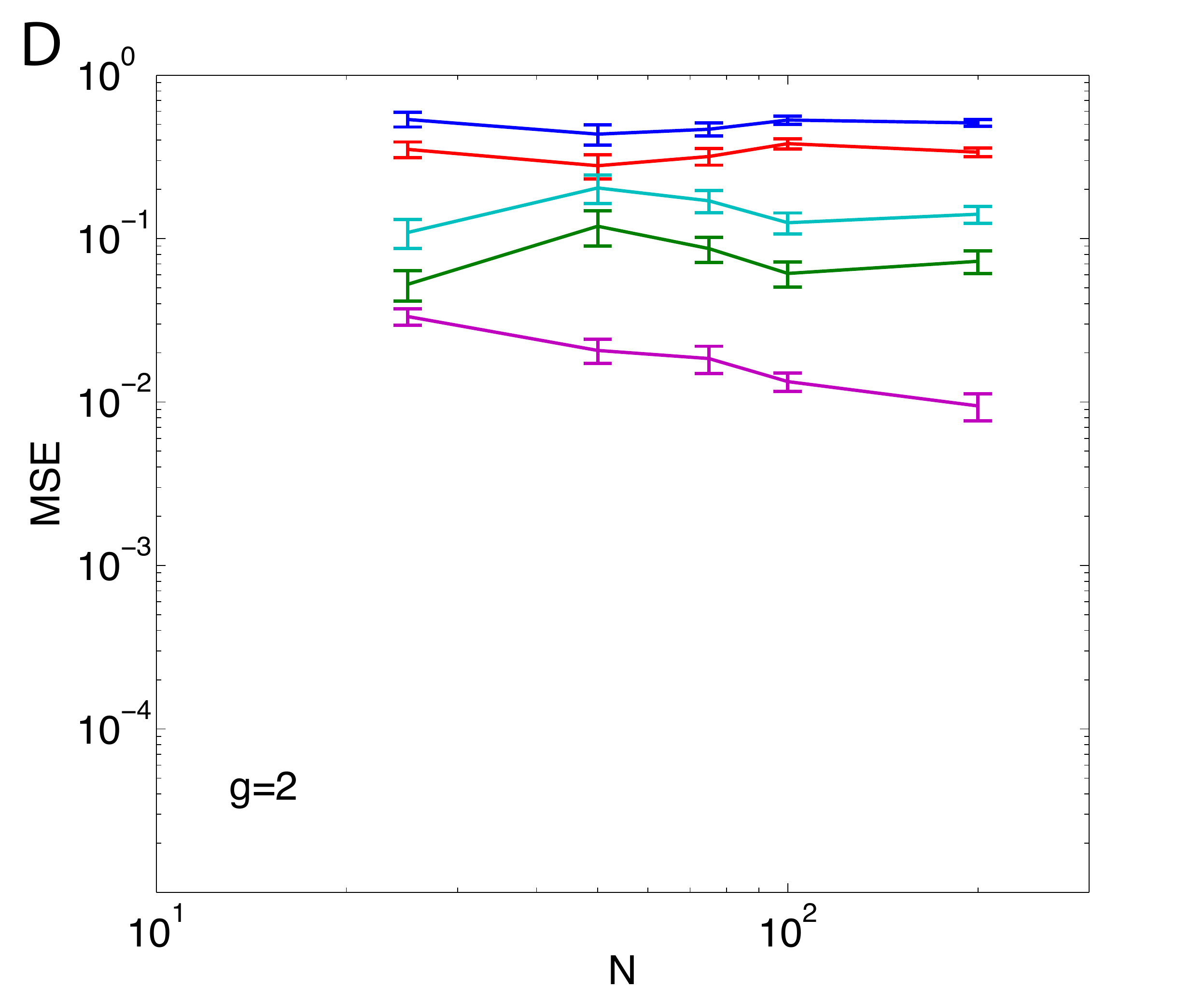}

\caption{Mean squared error of Naive Mean field (blue), RH-TAP (green), MS-MF (red), unconstrained Gaussian Average approach (light blue) and Extended Plefka (magenta) for predicting entire dynamics of magnetizations. The mean squared error is plotted as a function of the system size $N$. We have used 100 time steps and 50,000 repeats to calculate the experimental magnetizations and have averaged the errors over 10 realizations of the couplings. The error bars are standard deviations over these realizations. 
The number of sample trajectories used in the algorithm for the Extended Plefka method is $N_T=50000$ for $N=25,75,100$, while 
$N_T=100000$ for $N=200$.
Stationary external field drawn independently for each spin from a normal distribution (zero mean, standard deviation 0.5).  A: symmetric case $k=0$, couplings strength $g=1$. B: fully asymmetric case $k=1$, $g=1$. C: $k=0$, $g=0.1$; unconstrained Gaussian Average, MS-MF and RH-TAP curves overlap. D: $k=0$, $g=2$. Notice the different scale on the y-axis in C with respect to the other panels. }
\label{fig:SysSize}
\end{figure}

The scaling of the MSE errors with N is shown in Fig. \ref{fig:SysSize}. Numerical simulations show that the error of the Extended Plefka method decays with the system size $N$ for every value of the parameters $g$ and $k$, while the errors of the RH-TAP and  MS-MF approximations decrease with $N$ only in the range of the parameters for which the the approximations were developed, 
which corresponds respectively to a symmetric network with small couplings and to an asymmetric network. 
The error computed using Na\"{i}ve Mean field and unconstrained Gaussian Average approximations shows no scaling with $N$. 
This seems to suggest that the Extended Plefka expansion  provides an accurate mean field description of the dynamics. Notice however that evaluating the local moments with grater accuracy requires considering the whole history of the single spin trajectory and that the complexity of the algorithm described in section \ref{sec:ExtPlefka} scales with the degrees of freedom as $T^2 N + T N N_T$. 
To speed up the algorithm one could argue that, when the couplings  $J_{ij}$ scale as $1/\sqrt{N}$, the two sums 
 $$\sum_j J^2_{ij} C_j(t,t'); \quad \sum_j J_{ij}J_{ji} R_j(t,t')$$ appearing in (\ref{eq:noise_covariance}) and (\ref{eq:h}) can be replaced by their self-averaging value  $$g^2 \sum_j  C_j(t,t') ; \quad g^2 \frac {1-k^2}{1+k^2} \sum_j R_j(t,t'), $$where we considered the distribution (\ref{eq:Couplings2}) for the couplings. This would allow us to write a self-averaging version of (\ref{eq:resp_plefka}) and the computational cost of the algorithm would reduce to  $T^2 + T N N_T$. We postpone this analysis to future work.
 
\section{Summary and Discussion}

In this paper we studied new approximations for predicting the dynamics of the kinetic Ising model with arbitrary couplings. First we distinguished between the variational and field theoretical approaches to the Na\"{i}ve Mean Field theory for a generic Markov Chain and pointed out that the two do not coincide unless the transfer function is logistic, as is the case for the kinetic Ising model and there are no self-interactions in the system. (For an overview of the approaches to Na\"{i}ve Mean Field theory for the equilibrium case see \cite{opper2001advanced}.) For the specific case of the kinetic Ising model with discrete time parallel updates, we then proposed two approximations based on generating functional integral technique: the Gaussian Average Variational method and the Extended Plefka expansion. In the Gaussian Average Variational method we expand the generating functional of the process to first order around a high dimensional complex Gaussian integral and optimize the resulting expression. An unconstrained optimization of the parameters of this Gaussian function, in which we assume no structure for the covariance matrix, provides equations of motion which, as our numerical analysis indicates, perform at the naive mean field level for small couplings while they get close to RH-TAP equations \cite{roudi2011dynamical} for larger couplings. On the other hand making suitable assumptions on the covariance matrix, allows us to recover the MS-MF equations \cite{mezard2011exact}, known to be exact for fully asymmetric connectivities in the thermodynamic limit. 

Although we numerically compared the dynamics of magnetizations predicted from our dynamical equations with those of simulating the system, we did not study the relaxation dynamics that our dynamical equations predict for such systems analytically. Such an analysis has been performed in the case of the $p$-spin spherical spin glass model in \cite{biroli1999dynamical} where it is shown that the long term dynamics of the dynamical TAP equations for this system can be seen as descending though the free energy landscape.  For symmetric couplings and constant external fields, the synchronous update model that we have considered here in the long time will equilibrate to a Boltzmann distribution determined by the Peretto's Hamiltonian \cite{peretto1984collective}. The replica analysis for this model have not been performed and, therefore, we cannot make any statements as to what degree our extended Plefka  and variational equations will be in agreement with such analyses. However, we would like to note that for the asynchronous update Glauber dynamics, once stationary magnetizations are assumed, the RH-TAP equations will coincide with the standard static TAP equations \cite{thouless1977solution}. As noted before the equations derived here using the extended Plefka expansion are generalization of the RH-TAP equations \cite{roudi2011dynamical} and reduce to those if correlations and response functions are not taken into account. Static TAP equations -- which can be derived as the stationary limit of RH-TAP equations-- in turn, describe the multitude of local minima observed in the low temperature phase of the SK model, and whose consistency with the replica approach has been formally established \cite{cavagna2003formal}. The situation regarding the variational method is less clear, for one reason because, besides that little is known of the low temperature properties of the Peretto's Hamiltonian \cite{scharnagl1995relaxation}, the resulting dynamical equations can change with the ansatz chosen for the covariance matrix of the fields and conjugate fields. If no ansatz is assumed, out numerical results show that at low temperatures (strong couplings), the error in predicting the magnetizations approaches those of the RH-TAP equations, which as stated before lead to static TAP equations in the stationary state. We will leave it to future studies to explore this similarity and the relaxation dynamics predicted by the variational approach in more detail and analytically.

In the extended Plefka approach, by expanding the log generating functional in the coupling strength, while fixing first and second order moments over time, we approximate the true interacting dynamics by an effective single site dynamics. Namely, within the approximated description, each spin is subjected to an effective local field (\ref{eq:h}) that contains a retarded interaction with its own past values and a coloured Gaussian noise. The main difference with other mean field techniques is that the whole history of the single spin trajectory is taken into account in the equation for local order parameters.  Numerical simulations show that considering this term leads to greater accuracy in predicting local magnetizations for all values of couplings strength, coupling asymmetry and different choices of external fields. We find that this memory term is stronger for larger degree of symmetry of the network, and negligible when the couplings are uncorrelated: in this case the MS-MF approximation is retrieved.

The methods proposed in this paper are quite general in their scope and in theory can be used for studying the dynamics of other kinetic models. In particular, we find it interesting to see how these approximations perform for point process models from the Generalized Linear Model family, from which the kinetic Ising model is just one simple example. Furthermore, these methods can also be applied for inverse problems: inferring the interactions and fields given spin trajectories \cite{roudi2011mean}. In particular, given the fact that inference and learning in the presence of hidden nodes can be casted in a functional integral language \cite{dunn2013learning}, our methods can naturally lend themselves to developing novel approximations in this case for point processes. In fact, very recently, the extended Plefka approach has been used for learning and inference of the continuous variables in the presence of hidden nodes \cite{bravi2016inference}. 

\section*{Acknowledgments}
This work has been partially supported by the Marie Curie Initial Training Network NETADIS (FP7, grant 290038). YR and CB also acknowledge fundings from the Kavli Foundation and the Norwegian Research Council Centre of Excellence scheme. YR is also grateful to the Starr foundation for financing his membership at the IAS.  

\appendix

\section{Determinant of S}
\label{sec:DetS}
As was mentioned in section \ref{sec:Optimization} of the main text, in this appendix we demonstrate that the determinant of the matrix $\bS$ that appears in (\ref{eq:matrixS})-(\ref{eq:matrixS1}) equals one. We are going to prove it irrespectively to the specific details of matrices $\bgamma$ and $\blambda$ in (\ref{eq:matrixS1}).
Consider a complex matrix $\bS$ with the block structure defined in (\ref{eq:matrixS}), where $\blam (t,t+1)$ and $\bgam (t)$ are generic complex square matrices of order $N$.

In order to compute its determinant partition the matrix $\bS$ as follows:

\begin{equation}
\bS=
\left[
\begin{array}{c|ccccc}
\bS(0,0) & \bS(0,1)& 0 &0&0&\cdots\\
\hline
\bS(1,0) & \bS(1,1)& \bS(1,2) &0&0&\cdots\\
0&\bS(2,1) & \bS(2,2)& \bS(2,3) &0&\cdots\\
0 &0 & \bS(3,2)& \bS(3,3) &\bS(3,4)&\cdots\\
\cdots&\cdots&\cdots&\cdots&\cdots&\cdots
\end{array}
\right]
\label{eq:PartitionS}
\end{equation}

The determinant of this partitioned matrix can be formulated in terms of its blocks through the properties of Shur complements. Indeed for a generic matrix M:

\begin{equation}
M=
\left[
\begin{array}{cc}
A & B \\
C  & D \\
\end{array}
\right]
\quad
\det M=
\det A \det \left[ D-
C
A^{-1}
B
\right]
\label{eq:ShurCompl}
\end{equation} 

Since the square matrix $\bS(0,0)$ is invertible, as can be easily checked in (\ref{eq:matrixS1}), (\ref{eq:ShurCompl}) can be used to express the determinant of $\bS$ as:

\begin{equation}
\fl \det \bS =\det \bS(0,0) \det \left[ \underbrace{\bS^{\setminus 0}-
\left(
\begin{array}{c}
\bS(1,0) \\
0 \\
\vdots
\end{array}
\right)
\bS(0,0)^{-1}
\left(
\begin{array}{ccc}
\bS(1,0) & 0 & \dots
\end{array}
\right)
}_{\tilde{\bS}^{\setminus 0}}\right]
\label{eq:formulaDetS}  
\end{equation}
where we have denoted with $\bS^{\setminus 0}$ the bottom right matrix in the partition (\ref{eq:PartitionS}).

Notice that second term in $\tilde{\bS}^{\setminus 0}$ --- the Shur complement of $\bS(0,0)$ --- has the form:

\begin{equation}
\fl \left(
\begin{array}{c}
\bS(1,0) \\
0 \\
\vdots
\end{array}
\right)
\bS(0,0)^{-1}
\left(
\begin{array}{ccc}
\bS(1,0) & 0 & \dots
\end{array}
\right)=
\left[
\begin{array}{c|ccccc}
\hat{\bS}(1,1) & 0& 0 &0&0&\cdots\\
\hline
0 & 0 & 0 & 0 &0&\cdots\\
0 & 0 & 0 & 0 &0&\cdots\\
0 & 0 & 0 & 0 &0&\cdots\\
\cdots&\cdots&\cdots&\cdots&\cdots&\cdots
\end{array}
\right]
\end{equation}
with
\begin{equation}
\fl \hat{\bS}(1,1)=
\left[
\begin{array}{c|c}
0 & -i\mathbb{I} \\
\hline
-i\mathbb{I}  & \brho (1) \\
\end{array}
\right], \qquad \brho(1)=\blam(0,1)^\top \bgam(0)\blam(0,1)
\end{equation}
such that the matrix $\tilde{\bS}^{\setminus 0}$ in (\ref{eq:formulaDetS}) turns out having the same block form as $ \bS^{\setminus 0}$,

\begin{eqnarray}
\tilde{\bS}^{\setminus 0}&=&
\left[
\begin{array}{c|ccccc}
\tilde{\bS}(1,1) & \bS(1,2)& 0 &0&0&\cdots\\
\hline
\bS(2,1) & \bS(2,2)& \bS(2,3) &0&0&\cdots\\
0&\bS(3,2) & \bS(3,3)& \bS(3,4) &0&\cdots\\
0 &0 & \bS(4,3)& \bS(4,4) &\bS(4,5)&\cdots\\
\cdots&\cdots&\cdots&\cdots&\cdots&\cdots
\end{array}
\right]\\
\tilde{\bS}(t,t)&=&
\left[
\begin{array}{c|c}
0 & -i\mathbb{I} \\
\hline
-i\mathbb{I}  & \tilde{\bgam} (t) \\
\end{array}
\right]
\label{eq:Stilde}
\end{eqnarray}
and

\begin{equation}
\tilde{\bgam} (t)=\bgam (t)-\blam(t-1,t)^\top \tilde{\bgam}(t-1)\blam(t-1,t) \qquad .
\label{eq:GammaTilde}
\end{equation}

As a consequence $\tilde{\bS}^{\setminus 0}$ is a block tridiagonal matrix, just like $\bS$, and in order to compute its determinant one can apply (\ref{eq:formulaDetS}) again, to express $\det\tilde{\bS}^{\setminus 0}$ as a function of the determinant of $\tilde{\bS}(1,1)$. By repeatedly applying (\ref{eq:formulaDetS}) to the Shur complements $\tilde{\bS}^{\setminus t}$ of $\tilde{\bS}(t,t)$, one shows that the determinant of $\bS$ can be factorized into determinants of $\tilde{\bS}(t,t)$s. As proven for $t=0$ these matrices $\tilde{\bS}(t,t)$ preserve the structure of $\bS(t,t)$ and therefore their determinants are $1$. Finally:

\begin{equation}
\det \bS = \prod_t \det \tilde{\bS}(t,t)=1 
\end{equation}

\section{Inverse of S}
\label{sec:InvS}

In sections \ref{subsec:ExactSolution} and \ref{subsec:Consistency} we relate the optimal values of the variationl parameters and the magnetizations to the elements of the covariance matrix $\bS^{-1}$ in the framework of the Gaussian Average method. In this appendix we derive expressions for these elements, namely the correlations between field and conjugate fields $S_{2Nt+i,2Nt+i}=\langle g_i(t)^2\rangle_{L'_s}$, $S_{2Nt+N+i,2Nt+N+j}=\langle\hat{g}_i(t)\hat{g}_j(t)\rangle_{L'_s}$ and $S_{2Nt+i,2N(t+1)+N+j}=\langle g_i(t)\hat{g}_j(t+1)\rangle_{L'_s}$ --- where $\left\langle \cdot\right\rangle_{L'_s}$ indicates averages under the gaussian measure $e^{-L'_s}$, with $L'_s=\frac{1}{2}\cG \bS \cG$.

\subsection*{Variance $\langle g_i(t)^2\rangle_{L'_s}$}

Here we close the set of equations (\ref{eq:magn02}) for the magnetizations with equations for the variances $S_{2Nt+i,2Nt+i}$ in terms of the interaction matrix $\bS$, whose entries are linked to the magnetizations through (\ref{eq:matrixS})-(\ref{eq:lambda}). 

Recall that the inverse of the non-singular matrix $\bS$ can be computed as   \cite{friedberglinear}

\begin{equation}
\bS^{-1}_{ij}=\frac{(-1)^{i+j}\det\left( \left[ \bS\right]_{ji} \right) }{\det(\bS)}\label{eq:form0}
\end{equation}
where $\left[ \bS\right]_{ji}$ is the $ji$ minor of the matrix $\bS$, obtained removing the j-th row and the i-th column from the matrix itself. In case of the matrix defined by (\ref{eq:matrixS}), whose determinant equals $1$, the problem of inverting the matrix corresponds to computing the determinant of these minors. We now aim to calculate the determinant of $\left[ \bS\right]_{2Nt+i,2Nt+i}$ following the derivation of the determinant of $\bS$.

As in \ref{sec:DetS}, we start with factorizing out the determinant of the diagonal blocks $\bS(t',t')$ up to $\bS(t-1,t-1)$, according to (\ref{eq:formulaDetS}). Given that these all equal $1$, we can rewrite the determinant of $\left[ \bS\right]_{2Nt+i,2Nt+i}$ as:

\begin{equation}
 \det\left( \left[ \bS\right]_{2Nt+i,2Nt+i}\right) =\det\left(I(i,t)\right)
 \label{eq:inv0}
 \end{equation}
 where
 
\begin{equation} 
\fl I(i,t)\equiv\left[ \bS^{\setminus t-1}\right] _{ii}-\left(
\begin{array}{c}
\left[ \bS(t,t-1)\right]_{\setminus i}  \\
0 \\
\vdots
\end{array}
\right)
\tilde{\bS}(t-1,t-1)^{-1}
\left(
\begin{array}{ccc}
\left[ \bS(t-1,t)\right]^{\setminus i}  & 0 & \dots \\
\end{array}
\right)
\label{eq:inv01}
\end{equation}
and we have defined $\tilde{\bS}(t,t)$ in \ (\ref{eq:GammaTilde}). $\left[ \bS(t,t-1)\right]_{\setminus i} $ and $\left[ \bS(t,t-1)\right]^{\setminus i} $ have been obtained removing respectively the i-th row and the i-th column from  $\bS(t,t-1)$. $\left[ \bS^{\setminus t-1}\right] _{ii}$ is instead the $ii$ minor of $\bS^{\setminus t-1}$ defined analogously as $\bS^{\setminus 0}$ in \ref{sec:DetS}:

\begin{equation}
\fl \bS^{\setminus t-1}=
\left[
\begin{array}{ccccc}
\bS(t,t) & \bS(t,t+1)& 0 &0&\cdots\\
\bS(t+1,t) & \bS(t+1,t+1)& \bS(t+1,t+2) &0&\cdots\\
0&\bS(t+2,t+1) & \bS(t+2,t+2)& \bS(t+2,t+3) &\cdots\\
0 &0 & \bS(t+3,t+2)& \bS(t+3,t+3) &\cdots\\
\cdots&\cdots&\cdots&\cdots&\cdots
\end{array}
\right] .
\label{eq:matSminusT}
\end{equation}

One can easily see that $I(i,t)$ in (\ref{eq:inv0}) preserves the block form of $\left[ \bS^{\setminus t-1}\right] _{ii}$, namely

\begin{equation}
I(i,t)=
\left[
\begin{array}{c|ccccc}
\left[ \tilde{\bS}(t,t)\right]_{ii}  & \left[ \bS(t,t+1)\right]_{\setminus i} & 0 &0&0&\cdots\\
\hline
\left[ \bS(t+1,t)\right]^{\setminus i}  & &  &&&\\
 0& &  \bS^{\setminus t} &&&\\
 ... & & &&&
\end{array}
\right]
\label{eq:inv1}
\end{equation}
and therefore one can apply the formula in (\ref{eq:formulaDetS}) once more to factorize the determinant in (\ref{eq:inv0}) into a product of two determinants as follows:

\begin{equation}
\det\left( \left[ \bS\right]_{2Nt+i,2Nt+i}\right) =
\det\left( \left[ \tilde{\bS}(t,t)\right]_{ii} \right)\det\left(I(i,t+1)\right)
\label{eq:inv2}
\end{equation}
where $I(i,t+1)$ has been defined in (\ref{eq:inv01}).

With a bit of algebra it's possible to show that the matrix $I(i,t+1)$ in (\ref{eq:inv2}), has the very same structure as $\bS^{\setminus t}$ and consequently of $\bS$. Thus the second factor in the above equation is $1$ and what's left is to compute the determinant of the $ii$ minor of the matrix $\tilde{\bS}(t,t)$.

Given the structure of $\left[ \tilde{\bS}(t,t)\right]_{ii}$

\begin{equation}
\left[ \tilde{\bS}(t,t)\right]_{ii} \equiv 
\left[
\begin{array}{c|c}
0&-i\mathbb{I}_{\setminus i}\\
\hline
-i\mathbb{I}^{\setminus i}&\tilde{\bgam}(t)\\
\end{array}
\right] \qquad ,
\end{equation}
where $\tilde{\bgam}(t)$ has been defined in (\ref{eq:GammaTilde}), its determinant reduces to:

\begin{eqnarray}
\det\left( \left[ \tilde{\bS}(t,t)\right]_{ii} \right)&=&(-1)^{N-1} \det\left( \tilde{\bgam}(t)\right) \det\left(i\mathbb{I}_{\setminus i}\tilde{\bgam}(t)^{-1}i\mathbb{I}^{\setminus i}\right)  \nonumber\\
&=&(-1)^{2N-2} \det\left( \tilde{\bgam}(t)\right) \det\left(\left[ \tilde{\bgam}(t)^{-1}\right]_{ii} \right) \nonumber\\
&=&\tilde{\gamma}_{ii}(t)   \qquad .
\end{eqnarray}

Finally we will check that the diagonal elements of $\bS^{-1}$ we've just obtained are well defined variances by proving that they can take only positive values. In order to do that we will show that the matrix $\tilde{\bgamma}(t)$ is positive definite.  

By substituting $\bgamma$ and $\blambda$, using respectively (\ref{eq:gamma}) and (\ref{eq:lambda}), in (\ref{eq:invSrec}) one can express $\tilde{\bgamma}(t)$ in terms of $\tilde{\bgamma}(t-1)$, the matrix of the couplings $\bJ$ and the matrix $M_{ij}(t)\equiv \delta_{ij}\sqrt{(1-m_i(t)^2)}$ ($m$ are the magnetizations) as

\begin{equation}
\tilde{\bgamma}(t)=(\bJ \bM(t))(\bJ \bM(t))^\top + \bJ \bM^2(t)\tilde{\bgamma}(t-1)\bM^2(t)\bJ^\top
\label{eq:positiveGamma}
\end{equation}

The first matrix on the right hand side of (\ref{eq:positiveGamma}) is positive definite. Since the sum of two positive definite matrices is positive definite, it is left to show that the second term on the right hand side of (\ref{eq:positiveGamma}) is positive definite. We will prove it by induction. First of all given that $\tilde{\bgamma}(0)=\bgamma(0)$, from the definition of $\bgamma$ in (\ref{eq:gamma}), we know that $\tilde{\bgamma}(0)$ is positive definite. Then we assume that $\tilde{\bgamma}(t-1)$ is positive definite and we prove that $\bJ \bM^2(t)\tilde{\bgamma}(t-1) \bM^2(t)\bJ^\top$ is positive definite. If $\tilde{\bgamma}(t-1)$ is positive definite, it exist a matrix $A$ such that $\tilde{\bgamma}(t-1)=AA^\top$. Exploiting the latter one can rewrite:

\begin{equation}
\bJ \bM^2(t)\tilde{\bgamma}(t-1) \bM^2(t)\bJ^\top = (\bJ \bM^2(t)A)(\bJ \bM^2(t)A)^\top
\end{equation}

proving that the second term on the right hand side of (\ref{eq:positiveGamma}) is positive definite. Consequently $\tilde{\bgamma}(t)$ is a positive definite matrix and its diagonal entries take only positive values.

\subsection*{Correlations $\langle\hat{g}_i(t)\hat{g}_j(t)\rangle_{L'_s}$}

Here we will prove that the two point correlation function between conjugate fields $S_{2Nt+N+i,2Nt+N+j}$ is zero, as claimed in \ref{subsec:Consistency}, where it enters the proof of consistency of the optimal parameter $\hat{\eta}=0$. 

Similarly to the previous subsection we will use (\ref{eq:form0}) to invert the matrix $S$ and compute the determinant of the minor through Shur's complement formula (\ref{eq:ShurCompl}):

\begin{eqnarray}
\fl \bS^{-1}_{2Nt+N+i,2Nt+N+j}&=&\frac{(-1)^{4Nt+2N+i+j}\det\left( \left[ \bS\right]_{2Nt+N+i,2Nt+N+j} \right) }{\det(\bS)}\nonumber\\
\fl &=&(-1)^{i+j}\det\left(Y(i,j,t)\right)\label{eq:hatG1}
\end{eqnarray}
with
 
\begin{eqnarray} 
\fl Y(i,j,t)=\left[ \bS^{\setminus t-1}\right] _{N+i,N+j}-\nonumber\\
\fl \left(
\begin{array}{c}
\left[ \bS(t,t-1)\right]_{\setminus N+i}  \\
0 \\
\vdots
\end{array}
\right)
\tilde{\bS}(t-1,t-1)^{-1}
\left(
\begin{array}{ccc}
\left[ \bS(t-1,t)\right]^{\setminus N+j}  & 0 & \dots \\
\end{array}
\right)
\label{eq:hatG2}
\end{eqnarray}
and we have defined $\tilde{\bS}(t,t)$ in \ (\ref{eq:Stilde}) and $\bS^{\setminus t-1}$ in (\ref{eq:matSminusT}). $Y(i,j,t)$ in (\ref{eq:hatG2}) preserves the block form of $\left[ \bS^{\setminus t-1}\right] _{N+i,N+j}$, namely

\begin{equation}
\fl Y(i,j,t)=
\left[
\begin{array}{c|ccccc}
\left[ \tilde{\bS}(t,t)\right]_{N+i,N+j}  & \left[ \bS(t,t+1)\right]_{\setminus N+i} & 0 &0&0&\cdots\\
\hline
\left[ \bS(t+1,t)\right]^{\setminus N+j}  & &  &&&\\
 0& &  \bS^{\setminus t} &&&\\
 ... & & &&&
\end{array}
\right]
\label{eq:hatG3}
\end{equation}

Conversely to the previous section we cannot express the determinant of $Y(i,j,t)$ in terms of the Shur's complement of $\left[ \tilde{\bS}(t,t)\right]_{N+i,N+j}$, since the latter is a singular matrix. One has instead to resort to the Shur's complement of the matrix $\bS^{\setminus t}$ that we know is invertible and its determinant is $1$, having the same structure of the matrix $\bS$ (\ref{sec:DetS}):

\begin{eqnarray}
\fl \det Y(i,j,t) =
\det\bS^{\setminus t}\det\left(\left[ \tilde{\bS}(t,t)\right]_{N+i,N+j}-\nonumber\right.\\ 
\fl \left. \left(
\begin{array}{ccc}
\left[ \bS(t,t+1)\right]_{\setminus N+i}  & 0 & \dots \\
\end{array}
\right)
\left[\bS^{\setminus t}\right] ^{-1}
\left(
\begin{array}{c}
\left[ \bS(t+1,t)\right]^{\setminus N+j}  \\
0 \\
\vdots
\end{array}
\right)
\right)
\label{eq:hatG4}
\end{eqnarray}

Just like $\left[ \tilde{\bS}(t,t)\right]_{N+i,N+j}$, the matrix whose determinant is the second factor on the right-hand side of (\ref{eq:hatG4}) is singular: as can be easily checked its i-th column is null, regardless of the elements of $\left[ \bS^{\setminus t}\right] ^{-1}$. This completes the proof that $S_{2Nt+N+i,2Nt+N+j}=0$ for all $i,j=1,...,N$ and $t=0,...,T-1$.

\subsection*{Correlations $\langle g_i(t)\hat{g}_j(t+1)\rangle_{L'_s}$}

Here we will prove that the two point correlation function between conjugate fields $S_{2Nt+i,2N(t+1)+N+j}$ is zero, as claimed in \ref{subsec:Consistency}, where it enters the proof of consistency of the optimal parameter $\hat{\eta}=0$. The derivation is very similar to the one for $S_{2Nt+i,2Nt+N+j}$ in the previous subsection. 

We will use (\ref{eq:form0}) to invert the matrix $S$ and compute the determinant of the minor through Shur's complement formula (\ref{eq:ShurCompl}):

\begin{eqnarray}
\fl \bS^{-1}_{2Nt+i,2N(t+1)+N+j}&=&\frac{(-1)^{4Nt +3N+i+j}\det\left( \left[ \bS\right]_{2Nt+i,2N(t+1)+N+j} \right) }{\det(\bS)}\nonumber\\
\fl &=&(-1)^{3N+i+j}\det\left(Z(i,j,t)\right)\label{eq:hatGG1}
\end{eqnarray}
with
 
\begin{eqnarray} 
\fl Z(i,j,t)\equiv\left[ \bS^{\setminus t-1}\right] _{i,3N+j}-\nonumber\\
\fl \left(
\begin{array}{c}
\left[ \bS(t,t-1)\right]_{\setminus i}  \\
0 \\
\vdots
\end{array}
\right)
\tilde{\bS}(t-1,t-1)^{-1}
\left(
\begin{array}{ccc}
\left[ \bS(t-1,t)\right]  & 0^{\setminus N+j} & \dots \\
\end{array}
\right)
\label{eq:hatGG2}
\end{eqnarray}
and we have defined $\tilde{\bS}(t,t)$ in \ (\ref{eq:GammaTilde}) and $\bS^{\setminus t-1}$ in (\ref{eq:matSminusT}). $Z(i,j,t)$ in (\ref{eq:hatGG2}) preserves the block form of $\left[ \bS^{\setminus t-1}\right] _{i,3N+j}$, namely

\begin{equation}
\fl Z(i,j,t)=
\left[
\begin{array}{cc|ccc}
\left[ \tilde{\bS}(t,t)\right]_{\setminus i}  & \left[ \bS(t,t+1)\right]_{i, N+j} & 0 &0&\cdots\\
 \bS(t+1,t)  & \bS(t+1,t+1)^{\setminus N+j} &  \bS(t+1,t+2) &0&\cdots\\
 \hline
 0&\bS(t+2,t+1)&  &&\\
 ... & 0&\bS^{\setminus t+1} &&\\
 & ...&&&
\end{array}
\right]
\label{eq:hatG3}
\end{equation}

Analogously to the previous section we will now express the determinant of $Z(i,j,t)$ using the Shur's complement of the matrix $\bS^{\setminus t+1}$ that we know is invertible and its determinant is $1$, just like the matrix $\bS$, as shown in \ref{sec:DetS}:

\begin{eqnarray}
\fl \det Z(i,j,t) =
\det\bS^{\setminus t+1}\det\left(\left[
\begin{array}{cc}
\left[ \tilde{\bS}(t,t)\right]_{\setminus i}  & \left[ \bS(t,t+1)\right]_{i, N+j} \\
 \bS(t+1,t)  & \bS(t+1,t+1)^{\setminus N+j}
\end{array}
\right]\right. \nonumber\\
\fl \left.-\left(
\begin{array}{ccc}
0 &0&\cdots\\
\bS(t+1,t+2) &0&\cdots
\end{array}
\right)
\left[ \bS^{\setminus t+1}\right]^{-1}
\left(
 \begin{array}{cc}
 0&\bS(t+2,t+1)\\
 0 & 0\\
 ...& ...
\end{array}
\right)
 \right)
\label{eq:hatGG4}
\end{eqnarray}

The structure of $\left[ \bS^{\setminus t+1}\right]^{-1}$ reflects $\bS^{-1}$ structure:

\begin{equation}
\fl \left[ \bS^{\setminus t+1}\right]^{-1}=
\left[
\begin{array}{cccc}
\Omega(t+2,t+2) & \Omega(t+2,t+3)& \Omega(t+2,t+3) &\cdots\\
\Omega(t+3,t+2) & \Omega(t+1,t+1)& \cdots &\cdots\\
\Omega(t+4,t+2)&\cdots & \cdots& \cdots \\
\cdots&\cdots&\cdots&\cdots
\end{array}
\right] 
\label{eq:InvT+1}
\end{equation}
with
\begin{equation}
\Omega(t+2,t+2)=\left[
\begin{array}{cc}
\tilde{\gamma}(t+2) &\Delta\\
\Gamma & 0
\end{array}
\right]
\end{equation}
where $\Delta$ and $\Gamma$ are matrices of order $2N$. The block form of $\Omega(t+2,t+2)$ follows that of the diagonal blocks of $\bS^{-1}$, that was proven to be such in the previous sections of this appendix.

Using (\ref{eq:InvT+1}) one can check that the matrix whose determinant is the second factor on the right-hand side of (\ref{eq:hatGG4}) is singular: its (N+j)-th row is null. This completes the proof that $S_{2Nt+i,2N(t+1)+N+j}=0$ for all $i,j=1,...,N$ and $t=0,...,T-1$.

\section{Gaussian Average method: variance}
\label{sec:Variance}
In this Appendix we study the stability of the dynamical system for the matrix $\tilde{\bgamma}$ defined in the main text by (\ref{eq:invSrec}). In order to do that we first average (\ref{eq:invSrec}) over the distribution of the couplings introduced in section \ref{sec:NumResults} through (\ref{eq:Couplings1})-(\ref{eq:Couplings2}). Consider then entries of $\tilde{\bgamma}(1)$:

\begin{eqnarray}
\overline{\tilde{\gamma}_{ij}(1)}&=g^2(1+g^2\overline{\tilde{\gamma}_{lm}(0)} )\qquad &{\rm for }\qquad  k=0 \nonumber\\
\overline{\tilde{\gamma}_{ij}(1)}&=g^2/2(1+8g^2\overline{\tilde{\gamma}_{lm}(0)} )\qquad &{\rm for }\qquad  k=1
\label{eq:DynSys}
\end{eqnarray}
where the overbar indicates the average over the disorder, $k$ is the parameter controlling the asymmetry of the couplings, while $g$ is the coupling strength. Notice the different factors in (\ref{eq:DynSys}) due to the correlations between the couplings. If we consider the univariate analogous to (\ref{eq:DynSys}):

\begin{eqnarray}
x(t)&=g^2(1+g^2x(t-1) )\qquad &{\rm for } \qquad k=0\nonumber\\
x(t)&=g^2/2(1+8g^2x(t-1))\qquad &{\rm for } \qquad k=1
\label{eq:DynSysUni}
\end{eqnarray}
one can easily check that this dynamical system is characterized by a critical value for $g$, $g_0$ that discriminates between different stability classes for the system. Below $g_0$ $x$ in  (\ref{eq:DynSysUni}) converges to a finite value, while it grows exponentially in time for $g>g_0$. The critical value for this chaotic behavior is respectively $g_0=1$ for fully asymmetric couplings ($k=1$) and $0.7<g_0<0.8$ for fully symmetric ones ($k=0$) .

\begin{figure}[htp]

\centering

\begin{tabular}{cc}

\includegraphics[width=50mm,height=50mm]{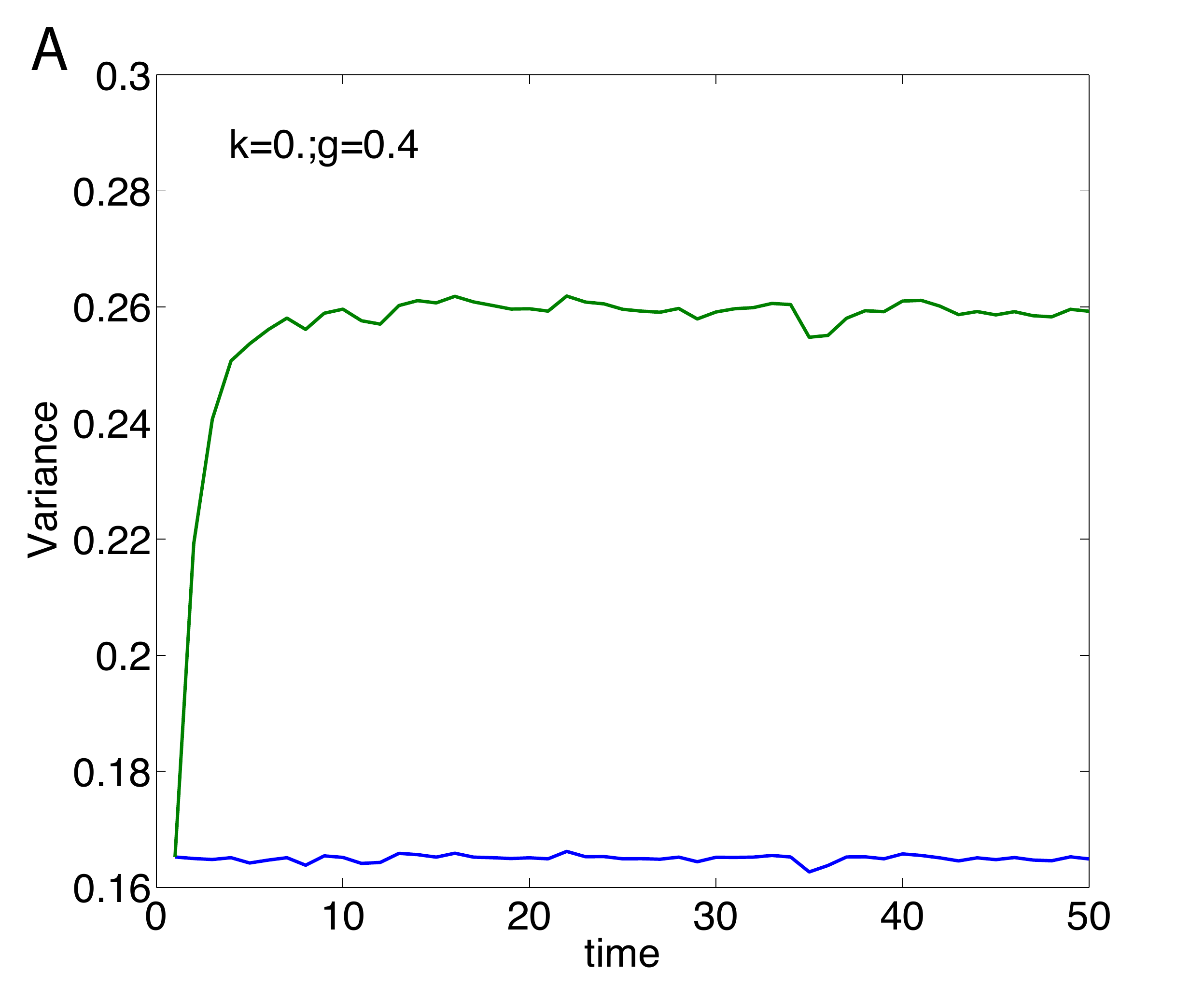}&

\includegraphics[width=50mm,height=50mm]{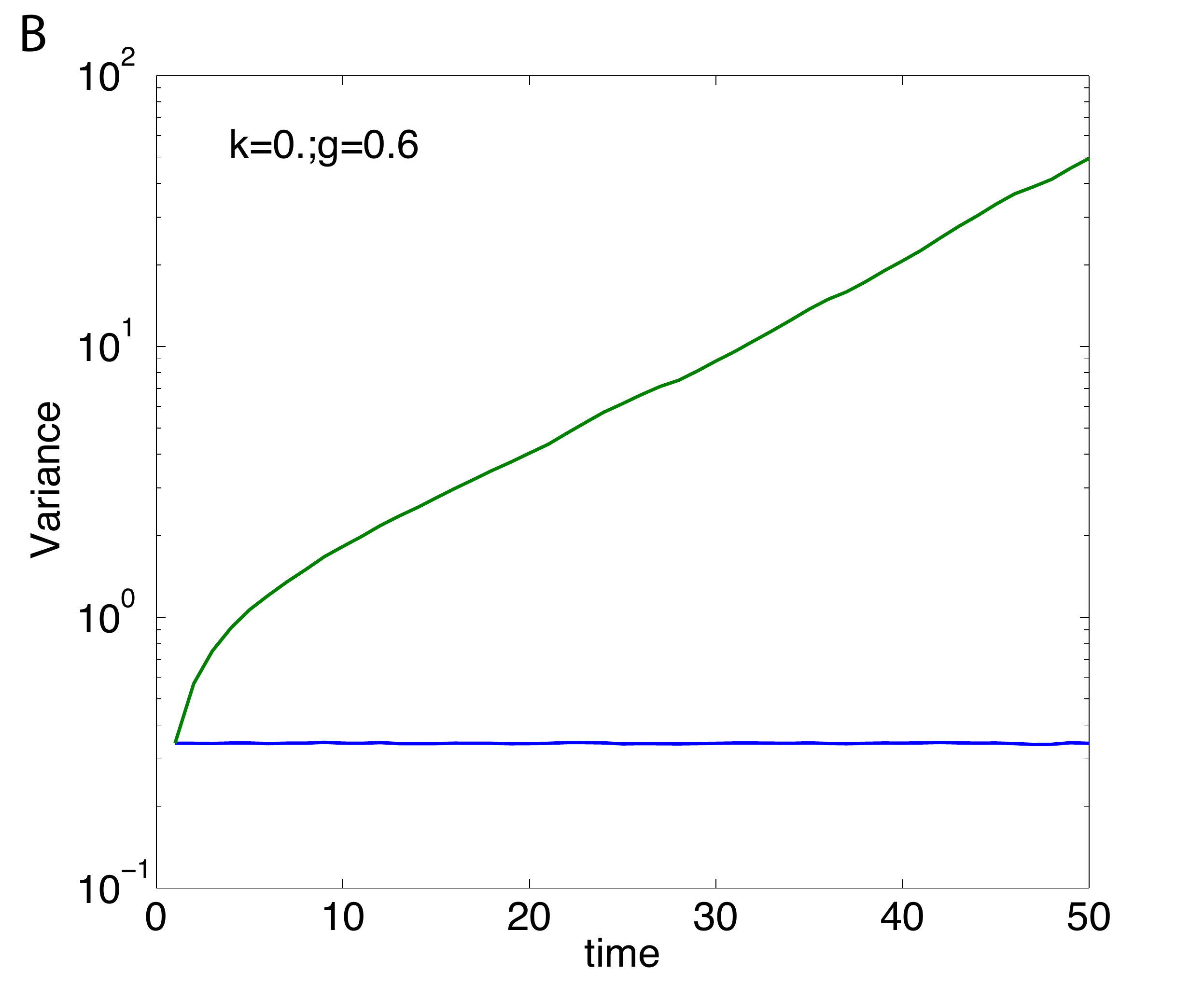}\\

\includegraphics[width=50mm,height=50mm]{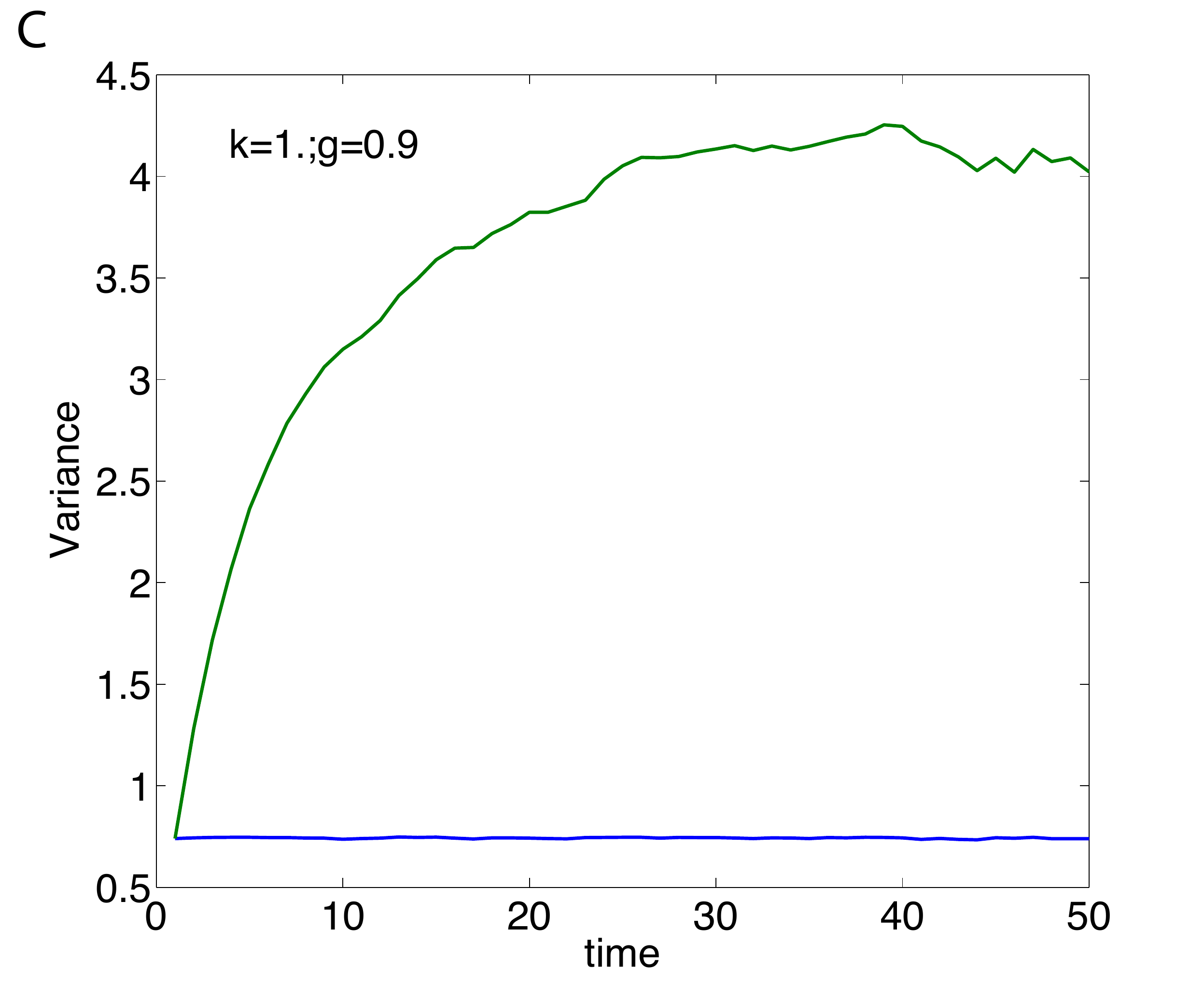}&

\includegraphics[width=50mm,height=50mm]{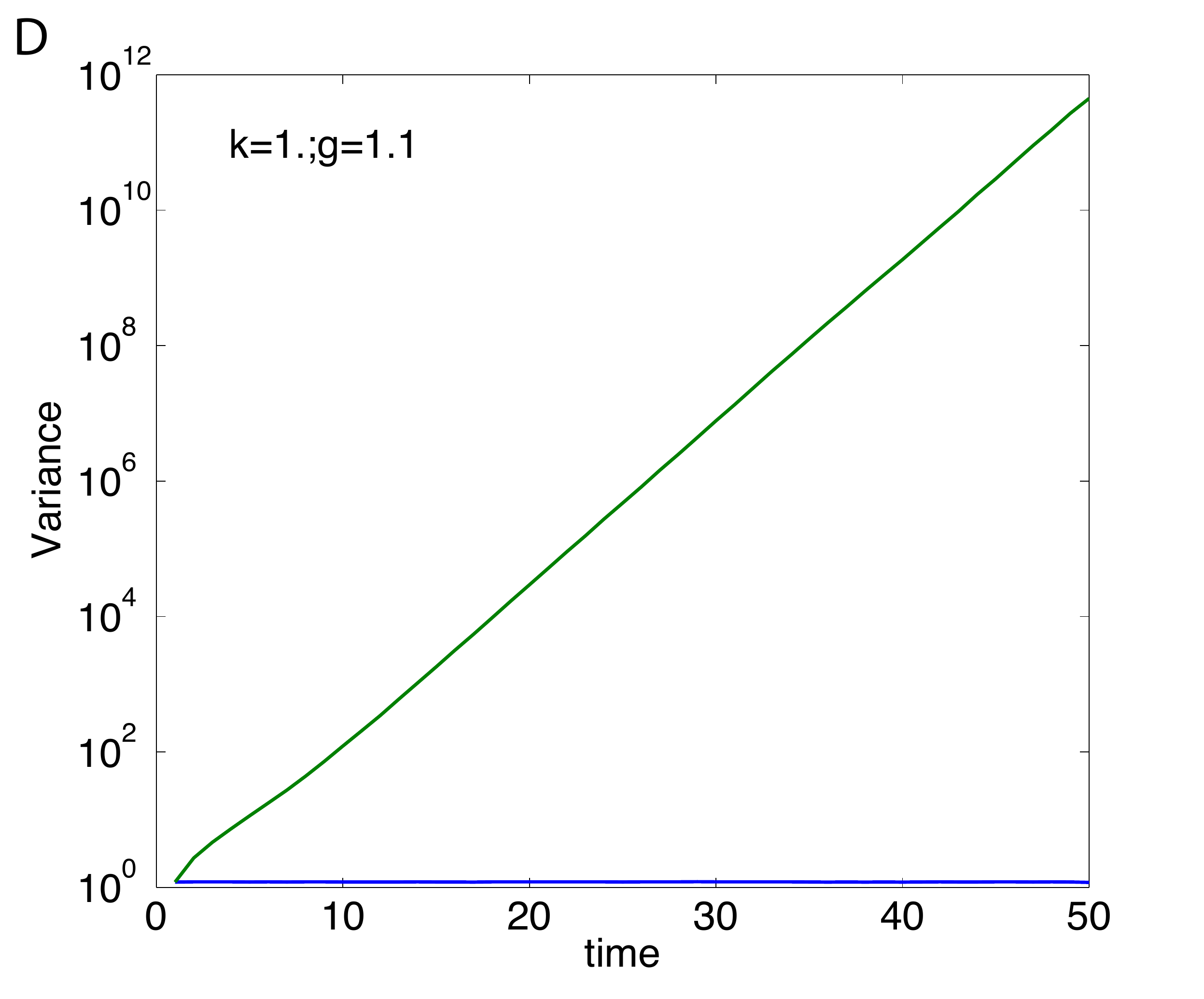}

\end{tabular}

\caption{The mean variance in the gaussian integral for the magnetizations versus time of reconstruction, when the experimental history of magnetization is known. Green: the Gaussian Average method; blue: the variance in \cite{mezard2011exact}. $N=20$, single realization of the couplings with $g=0.4$ (A),$g=0.6$ (B), $g=0.9$ (C) and $g=1.1$ (D) . The experimental magnetizations are computed using $10^4$ samples of the dynamics. Zero external field. Top: asymmetry parameter $k=0$. Bottom: asymmetry parameter $k=1$.  }

\label{Figure4}
\end{figure}

We got numerical evidence to support our intuition. We found that  the  variance in the Gaussian Average method undergoes a chaotic behavior when the couplings strength reaches a certain critical value, $g_0\sim 0.5$ for symmetric connectivities and $g_0\sim 1$ for fully asymmetric ones. This value depends on the degree of symmetry of the couplings, on the presence of the external field and there are small fluctuation across different realizations of the couplings, but the phenomenon is qualitatively conserved.  Figure \ref{Figure4} shows single realizations of the couplings below and above critical values and compares the mean of the Gaussian Average variances $\frac{1}{N}\sum_i\tilde{\gamma}_{ii}(t)$ with the mean of the MS-MF variances (mean of $\gamma_{ii}(t)$ (\ref{eq:gamma}) in our notation) in the gaussian integral for fully asymmetric couplings \cite{mezard2011exact}.

 \section{Details on the extended Plefka expansion.}
\label{sec:Plefka}

We rewrite the functional $\Gamma_{\alpha}$  (\ref{eq:Gamma_def}) 
as 
\begin{eqnarray}
\fl {\Gamma}_{\alpha}= \ln  \int d\mathcal{G} \; {\rm Tr}_s \exp(\Xi_{\alpha}[{\bold m},{\hat{\bold m}}, {\bold C},  {\bold B},{\bold R},{ \mathcal{G}} ,{\bold s}])
- N T \ln 2 \pi -N \ln 2,
\end{eqnarray}
where 
\begin{eqnarray}
\fl \Xi_{\alpha}[{\bold m},{\hat{\bold m}}, {\bold C},  {\bold B},{\bold R},{ \mathcal{G}} ,{\bold s}]=\sum_{i,t}   \left\{ i \hat{g}_i(t)
[ g_i(t) - \alpha \sum_j J_{ij}s_j(t) ] + s_i(t+1) g_i(t) 
 \right. \nonumber\\
\fl \left. - \ln 2 \cosh g_i(t)  -i h_i(t) [ \hat{g}_i(t)-\hat{m}_i(t)] +
 \psi_i(t) [ s_i(t)-m_i(t)] \right. \nonumber\\
\fl \left.+ \frac{1}{2} \sum_{t'} \hat{C}_i(t,t') [ s_i(t) s_i(t') -C_i(t,t')] 
 + \frac{1}{2} \sum_{t'} \hat{B}_i(t,t') [ \hat{g}_i(t) \hat{g}_i(t')- B_i(t,t') ] \right.\nonumber\\
\fl \left. -i \sum_{t'}\hat{R}_i(t',t) [\hat{g}_i(t) s_i(t')- i R_i(t',t) ]  \right\}, 
\label{eq:xi_funct}
\end{eqnarray}
and proceed with the perturbation expansion of $\Gamma_{\alpha}$ around $\alpha=0$ up to the second order:
\begin{equation} 
\Gamma_{\alpha} = \Gamma^{(0)} +\alpha \Gamma^{(1)} + \frac{\alpha^2}{2} \Gamma^{(2)},
\label{eq:Gamma_exp}
\end{equation}
where $\Gamma^{(k)}= {\partial^k \Gamma_{\alpha}}/{\partial \alpha^k}\vert_{\alpha=0}$.
At the end of the calculation we will set $\alpha=1$.
The first term in the expansion is given by
\begin{eqnarray}
\fl\Gamma^{(0)} [{\bold m},{\hat{\bold m}}, {\bold C},  {\bold B},{\bold R}]= \ln Z_0[{\boldsymbol \psi}^0, { \boldsymbol h}^0, \hat{\bold C}^0,\hat{\bold B}^0,\hat{\bold R}^0]
- \sum_{it}\psi^0_i(t)m_i(t) +i\sum_{it}h^{0}_i(t)\hat{m}_i(t) \nonumber\\
\fl-\frac{1}{2} \sum_{itt'} \hat{C}^0_i(t,t') C_i(t,t') - \frac{1}{2} \sum_{itt'}  \hat{B}_i^0(t,t') B_i(t,t')
 - \sum_{itt'}\hat{R}^0_i(t',t) R_i(t',t),
\label{eq:Gamma0}
\end{eqnarray}
where
\begin{eqnarray}
\fl Z_0[{\boldsymbol \psi}, { \boldsymbol h}, \hat{\bold C},\hat{\bold B},\hat{\bold R}]= \frac{1}{2^N(2\pi)^{NT}} \int d\mathcal{G}  \, {\rm Tr}_s \prod_{it} 
\exp \left\{\vphantom{\sum_{t'}^{t-1}} i \hat{g}_i(t) g_i(t) + s_i(t+1) g_i(t)  \right. \nonumber \\
\fl  \left. - \ln \cosh g_i(t) -i\hat{g}_i(t) h_i(t)+ \psi_i(t) s_i(t) + \frac{1}{2} \sum_{t'} \hat{C}_i(t,t') s_i(t) s_i(t')\right. \nonumber\\
\fl \left.   + \frac{1}{2} \sum_{t'} \hat{B}_i(t,t') \hat{g}_i(t) \hat{g}_i(t')
 -i \sum_{t'}\hat{R}_i(t',t) \hat{g}_i(t) s_i(t')  \right\} 
\label{eq:Z0_general}
\end{eqnarray}
and ${\boldsymbol \psi}^0, { \boldsymbol h}^0, \hat{\bold C}^0,\hat{\bold B}^0,\hat{\bold R}^0$ are the fields for 
which the set of equations (\ref{eq:param}) is satisfied for $\Gamma_{\alpha}=\Gamma^{(0)}$
for a given value of ${\bold m},{\hat{\bold m}}, {\bold C},  {\bold B},{\bold R}$.
We compute
$\Gamma^{(1)}$ as follows:
\begin{equation}
\Gamma^{(1)}=\left. \frac{\partial \Gamma_{\alpha}}{\partial \alpha}\right \vert_{\alpha=0}= \left. \left \langle \frac{\partial \Xi_{\alpha}}{\partial \alpha} 
 \right \rangle  \right \vert_{\alpha=0}.
\label{eq:G1}
\end{equation}
Using (\ref{eq:xi_funct}) one finds:
\begin{eqnarray}
\fl \frac{\partial \Xi_{\alpha}}{\partial \alpha} =
 -i \sum_{ijt} J_{ij} \hat{g}_i(t) s_j(t) 
-i \sum_{it}  \frac{\partial h_i(t)}{\partial \alpha}  [ \hat{g}_i(t)-\hat{m}_i(t)] +
 \sum_{it}  \frac{\partial \psi_i(t)}{\partial \alpha}  [ s_i(t)-m_i(t)] \nonumber\\
\fl  + \sum_{itt'} \frac{1}{2} \sum_{t'} \frac{\partial \hat{C}_i(t,t')}{\partial \alpha} [  s_i(t) s_i(t') -C_i(t,t')]
+\sum_{itt'} \frac{1}{2} \sum_{t'} \frac{\partial \hat{B}_i(t,t')}{\partial \alpha} [ \hat{g}_i(t) \hat{g}_i(t')- B_i(t,t') ] \nonumber \\
\fl -i\sum_{itt'} \sum_{t'}\frac{\partial \hat{R}_i(t',t)}{\partial \alpha} [ \hat{g}_i(t) s_i(t')- i R_i(t',t) ]. 
\label{eq:xi_deriv}
\end{eqnarray}
When computing the average $\langle {\partial \Xi_{\alpha}}/{\partial \alpha}  \rangle_{\alpha}$ as defined in (\ref{eq:average_alpha_def}),
all the terms on the right hand side of (\ref{eq:xi_deriv})
except for the first one vanish because of the set of equations (\ref{eq:moments}).
Moreover at $\alpha$=0 the spins are decoupled and the averages are trivial:
\begin{eqnarray}
\Gamma^{(1)}=-i \sum_{ijt} J_{ij} \langle  \hat{g}_i(t) s_j(t)\rangle_0 = -i\sum_{ijt} J_{ij} \hat{m}_i(t) m_j(t).
\label{eq:Gamma1}
\end{eqnarray}
For the second derivative of $\Gamma_{\alpha}$ with respect to $\alpha$ we have
\begin{equation}
 \frac{\partial^2 \Gamma_{\alpha}}{\partial \alpha^2}  =
\left \langle \frac{\partial^2 \Xi_{\alpha}}{\partial \alpha^2} 
 \right \rangle_{\alpha} +
\left \langle \left( \frac{\partial \Xi_{\alpha}}{\partial \alpha} \right)^2
 \right \rangle_{\alpha} -
\left( \left \langle \frac{\partial \Xi_{\alpha}}{\partial \alpha} 
 \right \rangle_{\alpha} \right)^2.
 \label{eq:Gamma_two}
\end{equation}
Using (\ref{eq:xi_deriv}) and  the set of equations (\ref{eq:moments}), it is easy to show that the first term on the right hand side of the above equation is zero.
One thus finds
\begin{equation}
\Gamma^{(2)}=\left. \left\langle \left(\frac{\partial \Xi_{\alpha}}{\partial \alpha} 
 -  \left\langle \frac{\partial \Xi_{\alpha}}{\partial \alpha}  \right\rangle  \right)^2 \right\rangle  \right \vert_{\alpha=0},
\label{eq:G2}
\end{equation}
which can be computed using (\ref{eq:xi_deriv}) and the following Maxwell equations:
\begin{eqnarray}
\left. \frac{\partial \psi_i(t)}{\partial \alpha} \right\vert_{\alpha=0} &=&
 -\frac{\partial}{\partial m_i(t)} \left. \frac{\partial \Gamma_{\alpha}}{\partial \alpha}\right\vert_{\alpha=0} =
i \sum_j J_{ji} \hat{m}_j(t) \nonumber\\
 i \left. \frac{\partial h_i(t)}{\partial \alpha}\right\vert_{\alpha=0}&=&
\frac{\partial}{\partial \hat{m}_i(t)} \left.\frac{\partial \Gamma_{\alpha}}{\partial \alpha} \right\vert_{\alpha=0}=
-i \sum_j J_{ij} m_j(t). 
\end{eqnarray}
Note that the derivatives of the two-time conjugate fields with respect to $\alpha$ are zero, e.g.
\begin{equation}
\frac{\partial \hat{C}(t,t')}{\partial \alpha}=\frac{\partial}{\partial C_i(t,t')}\frac{\partial \Gamma_{\alpha}}{\partial \alpha}=0.
\end{equation}
We finally obtain
\begin{eqnarray}
 \Gamma^{(2)}=-\sum_{iji'j'tt'} \langle \delta \hat{g}_i(t) J_{ij} \delta s_j(t) 
\delta \hat{g}_{i'}(t') J_{i'j'} \delta s_{j'}(t')\rangle_0,
\label{eq:Gamma_2_delta}
\end{eqnarray}
where we defined $\delta s_i(t)=s_i(t)-m_i(t)$ and $\delta \hat{g}_i(t)=\hat{g}_i(t) -\hat{m}_i(t)$.
Since the averages are taken at $\alpha=0$ spins at different sites are decoupled and the only 
 non-vanishing terms in (\ref{eq:Gamma_2_delta}) correspond to the case ${i'=i,\,j'=j}$ and ${i'=j,\,j'=i}$:
\begin{eqnarray}
\fl \Gamma^{(2)} = -\sum_{ijtt'} \left[ J_{ij}^2 \langle \delta \hat{g}_i(t) \delta \hat{g}_i(t') \rangle_0 \langle \delta s_j(t) \delta s_j(t') \rangle_0 \right. \nonumber\\
\left. +J_{ij}J_{ji} \langle \delta \hat{g}_i(t) \delta s_i(t') \rangle_0 \langle \hat{g}_j(t') \delta s_j(t) \rangle_0 \right],
\end{eqnarray}
which can be written in terms of the moments as follows
\begin{eqnarray}
\fl \Gamma^{(2)} =-\sum_{ijtt'} \left[ J_{ij}^2 \left(B_i(t,t') - \hat{m}_i(t) \hat{m}_i(t')\right)
 \left(C_j(t,t') - m_j(t) m_j(t')\right) \right. \nonumber\\
 \left. + J_{ij}J_{ji}\left(iR_i(t',t) - \hat{m}_i(t) m_i(t')\right)\left(iR_j(t,t') - \hat{m}_j(t') m_j(t)\right) \right].
\label{eq:Gamma2}
\end{eqnarray}

Inserting (\ref{eq:Gamma0}), (\ref{eq:Gamma1}) and (\ref{eq:Gamma2}) in  (\ref{eq:Gamma_exp})
we find the explicit expression of the functional $ \Gamma_{\alpha}$
expanded up to the second order. Considering the set of equations (\ref{eq:param}) within the second order expansion
and setting the auxiliary fields to zero, we can extract the value of the fields
${\boldsymbol \psi}^0, { \boldsymbol h}^0, \hat{\bold C}^0,\hat{\bold B}^0,\hat{\bold R}^0$
as functions of the correct (within the expansion) marginal first and second moments:

\begin{eqnarray}
\psi_i^0(t)&=& 0\nonumber\\
h^0_i(t)&=& h_i(t) + \sum_j J_{ij} m_j(t)  - \sum_{j t'} J_{ij}J_{ji} R_j(t,t') m_i(t') \nonumber\\
\hat{C}_{i}^0 (t,t')&=&0 \nonumber\\
\hat{B}_{i}^0 (t,t')&=& - \sum_{j}  J_{ji}^2  \left(C_j(t,t') - m_j(t) m_j(t')\right)\nonumber\\
\hat{R}_{i}^0 (t,t')&=&   \sum_{j} J_{ij}J_{ji}R_j(t',t).
\label{eq:aux_p}
\end{eqnarray}
From general results for generating functional analysis of spin systems
\cite{coolen2000}
it can be shown that $\hat{m}=0$, $B=0$ and  $R(t,t')$ 
 has the meaning of a local response function and is non-vanishing only for $t > t'$.
To get an explicit expression for $Z_0$ we insert (\ref{eq:aux_p}) in (\ref{eq:Z0_general}). 
It yields $Z^0[h]=\prod Z^0_i[h_i]$, where
\begin{eqnarray}
\fl Z_i^0[h_i] = \frac{1}{2^N(2\pi)^{NT}} \int d\mathcal{G}  \,  {\rm Tr}_{s_i}  \prod_{t} \exp \left\{ \vphantom{\sum_{t'}^{t-1}} s_i(t+1) g_i(t)
 - \ln \cosh g_i(t) -i\hat{g}_i(t) h_i(t)   \nonumber \right.\\
\fl \left. + i \hat{g}_i(t) g_i(t) -i \sum_j \hat{g}_i(t) J_{ij} m_j(t) - \frac{1}{2} \sum_{jt'} \hat{g}_i(t)  J_{ij}^2 
 \left[ C_j(t,t') - m_j(t) m_j(t') \right] \hat{g}_i(t')\right.\nonumber \\
\fl \left.  -i \sum_{jt'} \hat{g}_i(t) J_{ij}J_{ji} R_j(t',t) \left[ s_i(t') - m_i(t')\right]  \right\}.
\label{eq:Z0}
\end{eqnarray}
To linearize  the quadratic terms in ($\ref{eq:Z0}$), we introduce the Gaussian random 
variables  $\phi_i(t)$, independently for each $i$, with zero mean and 
covariance $\langle \phi_i(t) \phi_i(t') \rangle = \sum_j J^2_{ij} (C_j(t,t') - m_j(t)m_j(t'))$,  
 obtaining:
\begin{eqnarray}
\fl Z_i^0[h_i] = \frac{1}{2^N(2\pi)^{NT}} \left\langle \int d\mathcal{G}  \,  {\rm Tr}_{s_i}  \prod_{t} \exp \left\{ \vphantom{\sum_{t'}^{t-1}} s_i(t+1) g_i(t)
 - \ln \cosh g_i(t) + i \hat{g}_i(t) \left[\vphantom{\sum_{t'}^{t-1}} g_i(t) \right. \right. \right. \nonumber \\ 
\left. \left. \left. \fl 
 - \left(\phi_i(t)  + \sum_j J_{ij} m_j(t) 
 - \sum_{j} \sum_{t'}^{t-1} J_{ij}J_{ji} R_j(t,t') [s_i(t') - m_i(t') ] + h_i(t) \right) \right] 
 \right\}\right \rangle_{\phi_i}
\end{eqnarray}
From the above equation one can see that the moment $R_i(t,t')$ defined in (\ref{eq:moments})
can be written as an average over the fields $\phi_i(t)$ as follows

\begin{equation}
R_i(t,t')=\left\langle  \frac{\partial s_i(t)}{\partial  \phi_i(t')}  \right\rangle_{\phi_i},
\label{eq:respons}
\end{equation}
and can be interpreted as a response function.

\section{The Yule-Walker equations}
\label{sec:Walker}
We want to generate the Gaussian random field $\phi_i^k(t)$  for given trajectory $k$ and spin $i$ 
based on the past values of the field $\phi_i^k(0), \phi_i^k(1)\dots \phi_i^k(t-1)$.
Since the random variables $\phi_i^k(0), \phi_i^k(1)\dots \phi_i^k(t)$ are jointly  Gaussian distributed 
with zero mean,
we know that the conditional expectation $\widehat{ \phi}_i^k(t) \equiv E \{ \phi_i^k(t) \vert \phi_i^k(0), \phi_i^k(1)\dots \phi_i^k(t-1) \}$
is given by the linear estimate 
\begin{equation}
\widehat{ \phi}_i^k(t)= \sum_{r=0}^{t-1} a(r) \phi_i^k(r),
\label{eq:YuleWalker1}
\end{equation}
which also happens to be the best mean square estimate of  $\phi_i^k(t)$ 
 given $\phi_i^k(r), \; r=0 \dots t-1$.
The coefficients $a(0), a(1), \dots a(t-1)$ are such that the mean square value 
of the estimation error $E  \{  [ \phi_i^k(t)-\widehat{ \phi}_i^k(t)]^2 \}$ is minimum.
By the orthogonality principle, this condition holds if the following set of equations is satisfied
\begin{equation}
E \{  [ \phi_i^k(t) - \sum_{r=0}^{t-1} a(r) \phi_i^k(r  ) ] \phi_i^k(r')   \} =0, \quad r'=0 \dots t-1,
\end{equation}
which can be rewritten in matrix form as 
\begin{equation}
{\bold A} \mathcal{C} =\mathcal{ C}_t ,
\label{eq:YuleWalker2}
\end{equation}
where ${\bold A}=[a(0) \dots a(t-1)]$ is the vector of coefficients, $\mathcal{ C} $ is the correlation matrix with elements
$\mathcal{ C}(r,r')=E\{  \phi_i^k(r) \phi_i^k (r')   \}$ for $ r,r'=0 \dots t-1$
and $\mathcal{ C} _t$ is the vector with elements $\mathcal{C}_t(r)=E\{  \phi_i^k(t) \phi_i^k(r)   \}$ for $ r=0 \dots t-1  $. 
Since $\phi_i^k(t)-\widehat{ \phi}_i^k(t) \perp \phi_i^k(r)$ for every $r=0, \dots t-1$, from (\ref{eq:YuleWalker1}) we conclude that 
$\phi_i^k(t)-\widehat{ \phi}_i^k(t) \perp \widehat{ \phi}_i^k(t)$ and the error reduces to 
\begin{equation}
\fl E  \{  [ \phi_i^k(t)-\widehat{ \phi}_i^k(t)]^2 \}=  E  \{  [ \phi_i^k(t)-\widehat{ \phi}_i^k(t)]\phi_i^k(t) \}= 1 - \sum_j J_{ij}^2 m_j^2(t) - {\bold A} \mathcal{ C} _t.
\label{eq:YuleWalker3}
\end{equation}
Knowing that  $E\{  \phi_i^k(r) \phi_i^k(r')   \} = \sum_j J_{ij}^2 [C_j(r,r')-m_j(r)m_j(r')]\; r,r'=0 \dots t$, 
 we can compute the coefficients $a(r)$ from (\ref{eq:YuleWalker2}) and draw the Gaussian
random variable $\phi_i^k(t)$ with mean and covariance given, respectively, by (\ref{eq:YuleWalker1}) and (\ref{eq:YuleWalker3}).

\section*{References}
\bibliography{Bibliography}{}
\end{document}